\documentclass[a4paper,aps,pra,twocolumn,showpacs,preprintnumbers,amsmath,amssymb]{revtex4}
\usepackage{bbm}
\usepackage{amsmath}
\usepackage{verbatim}
\usepackage{graphicx}
\bibliography{References}
\begin{document}

\title{Efficient quantum memory and entanglement between light and an
atomic ensemble using magnetic fields}

\author{Christine A. Muschik$^1$, Klemens Hammerer$^1$, Eugene S. Polzik$^{2,3}$, J. Ignacio Cirac$^1$}

\affiliation{ $^1$Max-Planck--Institut f\"ur Quantenoptik,
Hans-Kopfermann-Strasse, D-85748 Garching, Germany \\
$^2$ QUANTOP, Danish Research Foundation Center for Quantum Optics, DK 2100 Copenhagen, Denmark\\
$^3$ Niels Bohr Institute, DK 2100 Copenhagen, Denmark}

\begin{abstract}
We present two protocols, one for the storage of light in an
atomic ensemble and the subsequent retrieval, and another one for
the generation of entanglement between light and atoms. They rely
on two passes of a single pulse through the ensemble, Larmor
precessing in an external field. Both protocols work
deterministically and the relevant figures of merit - such as the
fidelity or the EPR variance - scale exponentially in the coupling
strength. We solve the corresponding Maxwell-Bloch equations
describing the scattering process and determine the resulting
input-output relations which only involve one relevant light mode
that, in turn, can be easily accessed experimentally.
\end{abstract}

\pacs{03.67.Mn, 32.80.Qk}

\maketitle 

\section{ Introduction}
Recent years have seen significant progress towards an efficient
quantum interface between light pulses carrying quantum
information and atomic ensembles suitable for storing and
processing this information. Two approaches based on probabilistic
photon detection \cite{CRFPEK,CMJLKK,EAMFZL} or on deterministic
homodyne measurements \cite{JSCFP,BP} have been developed. Of
particular importance in the context of quantum information are
means to swap the state of light and atoms - enabling a quantum
memory for light - and to create Einstein-Podolsky-Rosen (EPR)
type of entanglement of light and atoms - the basic resource for
quantum teleportation.

Concerning the quest for a quantum memory, an important
experimental advance was the recent demonstration of the storage
of weak coherent light pulses in atoms \cite{JSCFP}, based on a
Quantum Non Demolition (QND) interaction, measurement of light and
feedback on atoms. However, reliable retrieval of the stored state
by means of the same protocol would require the use of short
pulses of squeezed light which are difficult to couple to atomic
ensembles in an efficient way. The design of less demanding
protocols for storage and retrieval of states of light remained a
challenge, also from a theoretical perspective. Several protocols
have been put forward, all relying on multiple passes of light
through the atomic ensemble \cite{KP,KHGC,HMPC,F,SSFMP,FSOP}. The
most efficient of these schemes, complying with the experimental
requirement to use Larmor precessing atomic spins, require eight
passes of a single pulse \cite{SSFMP} or two pulses each crossing
twice an atomic cell \cite{FSOP}. In this paper we present a
protocol, which consists of only \textit{two} passes of a
\textit{single} pulse and achieves a state exchange of light and
atoms scaling \textit{exponentially} in the coupling strength
$\kappa$, defined operationally as the signal to noise ratio of
the underlying QND interaction. This scheme allows one to perform
the complete transfer of a quantum state of light onto atoms and
back under modest experimental conditions, as we show for both,
coherent states as well as arbitrary superpositions of vacuum and
a single photon Fock state.

Moreover, the same double pass setup serves with a slightly
changed geometry as a deterministic source of EPR entanglement
between light and atoms. The entanglement scales thereby again
exponentially in $\kappa$. Together, these two protocols add to
the growing toolbox for quantum information processing with room
temperature atomic vapors, which has already provided the
possibility to entangle two atomic ensembles via a
Bell-measurement on two Larmor precessing spins \cite{JKP}. In
combination these tools undoubtedly pave the way towards numerous
relevant applications, of which the demonstration of a complete
quantum memory and quantum teleportation are just the most
immediate.

To be more specific, the setup of both protocols consists of an
ensemble at room temperature in a cubic glass cell. It is placed
in an external magnetic field with large spin polarization along
the axis of this field, such that the transverse spin components
precess at frequency $\Omega$. A coherent pulse is directed
through the atomic sample such, that it crosses it twice under an
angle of 90 degrees in the plane orthogonal to the axis of the
magnetic field. The length $d$ of the loop in the optical path is
small, such that Larmor precession is frozen on a time scale
$d/c\ll\Omega^{-1}$, but the pulse length is large as compared to
the Larmor period, $T\gg\Omega^{-1}$. Under these conditions and
the assumption that $\Omega T\gg\kappa^2$, which is well fulfilled
in current experiments, we carefully solve the Maxwell-Bloch
equations describing the dynamics of this scattering process. We
identify the relevant light modes, which can be stored and
retrieved or get entangled with atoms and characterize their
temporal profile. The central frequency of these modes lies at the
upper or lower sideband of the carrier frequency, which is to be
expected given the splitting of ground state levels of $\Omega$,
and their slowly varying amplitude is exponential of the form
$\exp(\pm\kappa^2t/2T)$. The modes can thus be easily accessed.
Note that this setup is, apart from the magnetic field, similar to
the one treated in \cite{SSFMP}. It is precisely the presence of
the magnetic field what enables us to achieve our results with the
simple setup described above.

The rest of the paper is organized as follows. In section
\ref{BasicIdeaCentralResults} we introduce the basic idea of our
protocol and summarize the central results. In section
\ref{Quantum memory} and \ref{Two mode squeezing} we provide the
detailed derivation for the quantum memory and the EPR source
respectively. These sections are supplemented by two appendices.
Finally, section \ref{Consideration of Noise} deals with sources
of noise under realistic conditions.
\section{Basic idea and central results}\label{BasicIdeaCentralResults}
\begin{figure}[pbt]
\begin{center}
\includegraphics[width=4cm]{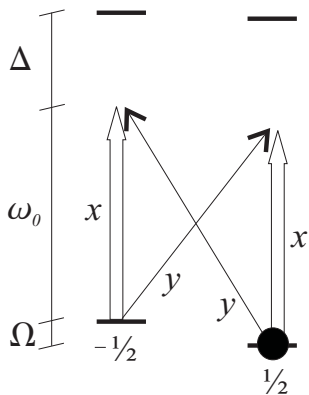}
\caption{Relevant internal levels with quantization along
$\hat{x}$. Thick arrows represent the strong coherent field in
$\hat{x}$ polarization, thin arrows indicate the quantum field in
$\hat{y}$ polarization. \label{Levelscheme1}}
\end{center}
\end{figure}

\begin{figure}[pbt]
\begin{center}
\includegraphics[width=6cm]{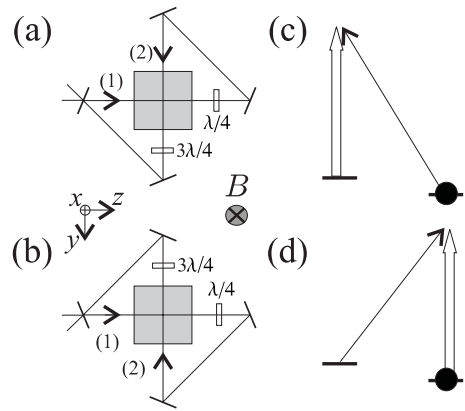}
\caption{Setups for having (a) a beam splitter or (b) a two mode
squeezing like dynamics. (c) and (d) show the effective
transitions.}. \label{cavity}
\end{center}
\end{figure}

\begin{figure}[pbt]
\begin{center}
\includegraphics[width=8cm]{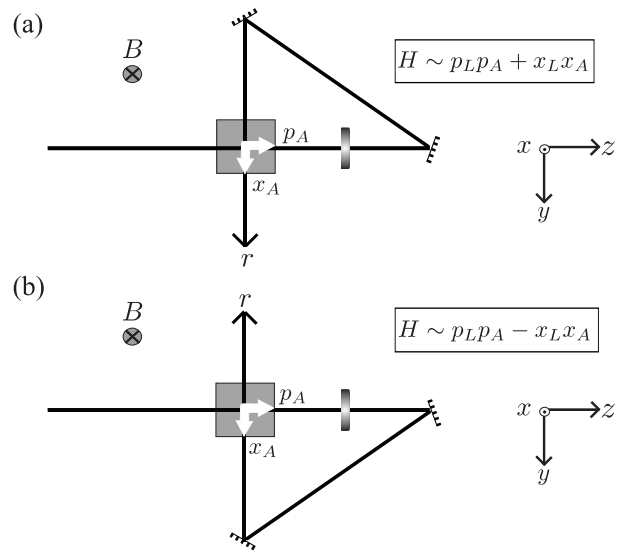} \caption{Schemes for
realization of a quantum memory (a) and a source of EPR
entanglement (b). \textbf{(a)} In the first pass a
$p_{L}p_{A}$-interaction occurs. Subsequently the pulse is sent
through a $\frac{\lambda}{4}\ $- plate, which interchanges $x_{L}$
and $p_{L}$. The pulse is reflected back onto the sample. This
happens at a timescale much shorter than the Larmor precession of
the atoms. Therefore the transverse components of the collective
spin can be assumed to remain in their place to a very good
approximation. Finally the pulse passes the atoms along $\hat{y}$.
Due to the changed geometry the atomic quadratures are also
interchanged. $p_{A}\rightarrow x_{A}$ and $x_{A}\rightarrow
-p_{A}$, which means, that the light field couples to $x_{A}$ in
it's second passage, hence leading to a $x_{L}x_{A}$ -interaction.
In \textbf{(b)} the changed geometry introduces a different sign
in the exchange of atomic quadratures, which leads to a
$-x_{L}x_{A}$ interaction in the second pass.\label{Setups}}
\end{center}
\end{figure}

In the following we consider a cubic atomic ensemble at room
temperature, which is placed in a magnetic field and interacts
with a pulse of light propagating along $\hat{z}$. The atomic
sample is assumed to be spin polarized along $\hat{x}$, while the
magnetic field is orientated along the opposite direction. The
pulse of light consists of a strong coherent $\hat{x}$-polarized
component of central frequency $\omega_{0}$, which is detuned by
$\Delta$ from the atomic transition, and a copropagating quantum
field in $\hat{y}$ polarization. The beam's waist is assumed to
cover most of the samples cross section. Atoms have a relevant
internal structure as shown in figure \ref{Levelscheme1}.
With $\hat{x}$ being the quantization axis, the classical light
field drives the $m=\pm 1/2\rightarrow m'=\pm1/2$ transitions, while
the copropagating quantum field couples to $m=\mp 1/2\rightarrow
m'=\pm1/2$. In the case of a dominant ground state population of
$m=1/2$ levels, creation and annihilation operators of collective
atomic excitations can be defined by
$b^{\dag}=\Sigma_{i}|-1/2\rangle\langle1/2|/ \sqrt{N_{A}}$ and $b$,
respectively, where $N_{A}$ is the total number of atoms in the
ensemble. Creation of an atomic excitation will then be accompanied
by the absorption (emission) of a photon at frequency
$\omega_0+\Omega,\,(\omega_0-\Omega)$, that is, at the upper (lower)
sideband, where $\Omega$ is the Larmor frequency. Note that only the
polarization, and not the energy of the sideband photons are
relevant, so the notion of upper/lower sideband is rather arbitrary.
Although we will finally deal with light interacting with atoms in
free space, it is instructive to consider first the case, where
atoms are placed inside a cavity supporting both sideband modes.
Related setups employing cavities are considered in \cite{PSC,
DCPG}. We assume in the following that the cavity life time is
much smaller than the Larmor period $\Omega^{-1}$ and let the
creation operators for the upper and lower sideband be given by
$a^{\dag}_{us}$ and $a^{\dag}_{ls}$ respectively. In the
dispersive limit, the effective Hamiltonian describing the
interaction is given by \mbox{$H\propto(b\,
a_{us}^{\dag}-b^{\dag}a^{\dag}_{ls}+\mathrm{h.c.})$}, where the
signs follow from Clebsch-Gordan coefficients. Note that if ground
state levels were degenerate, such that $a_{us}=a_{ls}\equiv a$,
the Hamiltonian would be \mbox{$H\propto (b-b^\dag)(a-a^\dag)$},
which is well known from the theory of quantum non-demolition
(QND) measurements of atomic spins. Including Zeeman splitting,
the interaction consists of a passive and an active part, $
H=H_{pas}-H_{act}$, where the passive part is a beam splitter
Hamiltonian $H_{pas}\propto b\ a_{us}^{\dag}+\mathrm{h.c.}$ and
acts only on the upper sideband, while the active part
$H_{act}\propto b^{\dag}a^{\dag}_{ls}+h.c.$ can be identified with
a two-mode-squeezing interaction, which involves exclusively the
lower sideband. Now, either of these two interactions can be
selected in one of the setups shown in figure \ref{cavity}a or
\ref{cavity}b. The interaction in every second pass will again be
given by $H$ but with phase changes $a_{ls(us)}\rightarrow
ia_{ls(us)}$, due to the $\lambda/4$ wave plate, and $b\rightarrow
\pm ib$, due to the change of the direction of light propagation,
where the upper sign holds for setup in figure \ref{cavity}a and
the lower for \ref{cavity}b. The resulting Hamiltonian is
\mbox{$H'_\pm\propto\pm(b\, a_{us}^{\dag}+
b^{\dag}a^{\dag}_{ls}+\mathrm{h.c.})$}. Together, we get for setup
\ref{cavity}a an interaction $H+H'_+=H_{pas}$ and for
\ref{cavity}b $H+H'_-=-H_{act}$. In either setup one of the two
$\Lambda$-type transitions in figure \ref{Levelscheme1} is
canceled by interference, and one is left with the transitions
shown in figures \ref{cavity}c and \ref{cavity}d. Note that these
configurations remind of the Raman scattering processes put
forward in \cite{DCLZ} for the realization of a quantum repeater.
Without a cavity, in setups as shown in figure \ref{Setups}, the
effects still persists, as we will show by solving the corresponding
Maxwell-Bloch equations. In contrast to the dynamics inside a
cavity, where Larmor precession of the atomic spin is not crucial,
it is well so for propagation in free space. This can be understood
by noting that both setups shown in figure \ref{Setups} possess a
certain asymmetry in how the two transverse spin components in
$\hat{y}$ and $\hat{z}$ direction are affected by light. This was
calculated in great detail in \cite{SSFMP}, where amongst others the
setup of figure \ref{Setups} was examined without magnetic field. We
emphasize that Larmor precession helps to remove this asymmetry.

In the rest of this section we collect the results for both, the
quantum memory and the two mode squeezing protocol. This will be
done in the language of canonical operators
$x_A=(b+b^\dag)/\sqrt{2}$ and $p_A=-i(b-b^\dag)/\sqrt{2}$ and
likewise for light, since solutions to Maxwell Bloch equations are
more conveniently derived in this formalism.
\paragraph*{Quantum memory} Within the memory scheme, figure \ref{Setups}a, the transfer of a quantum state of
light onto atoms or vice versa approaches perfect mapping
exponentially in the coupling strength. We have
\begin{equation*}
\left(%
\begin{array}{c}
  x^{out}_{A} \\
  p^{out}_{A} \\
\end{array}%
\right)=e^{-\frac{\kappa^{2}}{2}}\left(%
\begin{array}{c}
  x^{in}_{A} \\
  p^{in}_{A} \\
\end{array}%
\right)+ \sqrt{1-e^{-\kappa^{2}}}\left(%
\begin{array}{c}
  x_{L +}^{in}  \\
  p_{L +}^{in}  \\
\end{array}%
\right),
\end{equation*}
for the write-in procedure, where $x_{A}^{in/out}$ and
$p_{A}^{in/out}$ are the atomic input/output quadratures of the
scheme and $x_{L+}^{in/out}$ and $p_{L+}^{in/out}$ refer to the
write-in light mode. It lies at the upper sideband (according to the
configuration considered above) and is modulated by a slowly varying
envelope with an exponential profile, which is a propagation effect.
For the retrieval the inverse accented light mode
$\acute{x}_{L-}^{in/out}$ and $\acute{p}_{L-}^{in/out}$ is used and
we have
\begin{equation*}
 \left(%
  \begin{array}{c}
   \acute{x}^{out}_{L-} \\
   \acute{p}^{out}_{L-}  \\
   \end{array}%
 \right)=-\sqrt{1-e^{-\kappa^{2}}}
  \left(%
  \begin{array}{c}
   x^{in}_{A} \\
   p^{in}_{A}  \\
   \end{array}%
 \right)
 +e^{-\frac{\kappa^{2}}{2}}
 \left(%
  \begin{array}{c}
   \acute{x}^{in}_{L+} \\
   \acute{p}^{in}_{L+}  \\
   \end{array}%
 \right),
\end{equation*}
where accents indicate quadratures referring to the read-out pulse.
Note that for large $\kappa$ the state exchange is perfect. It is
remarkable that both pairs of input-output relations have a form
which reminds of a decoherence process, with the important
difference that we have modes in place of Langevin noise operators,
which can be controlled at will. The fidelity for the complete state
transfer - write in and subsequent retrieval of a state of light- is
given in figure \ref{Fidelities}(a) and (b) for coherent input
states and light qubits respectively.

\paragraph*{EPR source} The active version of the protocol, figure \ref{Setups}b, generates correlations between
atoms and light, which grow exponentially in the coupling. One can
define interspecies EPR modes
\begin{align*}
x_{1} &=\frac{1}{\sqrt{2}}(x_{A}-\tilde{p}_{L+}),& p_{1} &=
\frac{1}{\sqrt{2}}(p_{A}+\tilde{x}_{L+}),\\
x_{2}&=\frac{1}{\sqrt{2}}(x_{A}+\tilde{p}_{L+}),& p_{2} &=
\frac{1}{\sqrt{2}}(p_{A}-\tilde{x}_{L+}).
\end{align*}
where $\tilde{x}_{L+}$, $\tilde{p}_{L+}$ refer to a light mode,
which resembles the mode $x_{L+}$, $p_{L+}$ introduced above apart
from the fact that the lower sideband is involved instead of the
upper one. $x_{1}$ and $p_{2}$ are squeezed, while $x_{2}$ and
$p_{1}$ are antisqueezed,
\begin{eqnarray*}
(\Delta x_{1})^{2}&=&(\Delta
p_{2})^{2}=e^{-2z},\\
(\Delta p_{1})^{2}&=&(\Delta x_{2})^{2}=e^{2z},
\end{eqnarray*}
where $z=\cosh^{-1}(e^{\frac{\kappa^{2}}{2}})$. The EPR variance
of
the generated state is depicted in figure \ref{Entanglement}.\\
The results presented above will be derived in the following
sections. We remark that each protocol can be realized involving
either the upper or the lower sideband. The sideband mode involved
can be changed by either inverting the ground state polarization
or changing the orientation of the magnetic field. Losses will be
considered in section \ref{Consideration of Noise} and it will be
shown that the proposed protocols are robust against the dominant
sources of noise.
%
%
\section{Quantum memory}\label{Quantum memory}
The following calculation will be done by means of canonical
operators. For atoms canonical variables are defined by means of the
Holstein-Primakoff transformation and approximation
\cite{Holstein-Primakoff}. Via the Holstein-Primakoff transformation
spin-eigenstates are mapped onto harmonic oscillator-eigenstates.
The initial coherent atomic spin state is treated as harmonic
oscillator ground state. The Holstein-Primakoff approximation allows
one to define the canonical atomic variables $x_{A}$ and $p_{A}$
corresponding to the $\hat{y}$ and $\hat{z}$ component of the
collective angular momentum $J$, $x_{A}=J_{y}/ \sqrt{\langle
J_{x}\rangle}$ and $p_{A}=J_{z}/ \sqrt{\langle J_{x}\rangle}$.
The light field in $\hat{y}$-polarization is described by
spatially localized modes
\begin{eqnarray}\label{Light modes}
x_{L}(r)&=&\frac{1}{\sqrt{4 \pi }}\int_{b}d\omega(a_{\omega}e^{
 -i (\omega_{0}-\omega) r/c}+h.c.),\nonumber\\
 p_{L}(r)&=&-\frac{i}{\sqrt{4 \pi }}\int_{b}d\omega(a_{\omega}e^{
 -i (\omega_{0}-\omega) r/c}-h.c.),
\end{eqnarray}
where the range of integration $b$ is a small bandwidth around the
carrier frequency $\omega_{0}$ containing $\Omega$. The spatial
argument $r$ refers to the distance along the optical path shown in
figure \ref{Setups} and we have $[x_{L}(r),p_{L}(r')]=ic\delta
(r-r')$, where $c$ is the speed of light and the width of the delta
function is on the order of $c/b$
Within this description the Hamiltonian for the off-resonant
scattering interaction takes the form $H\propto p_{L}p_{A} $
\cite{DCZP}. Detailed descriptions can be found in
\cite{Teleportation} and \cite{J}.
%
%
\subsection{Write-in}\label{Write-in}
The double-pass interaction in setup \ref{Setups}a can be
described by
 \begin{eqnarray*}
 H=H_{atoms}+H_{light}+V_{1}+V_{2}\ .
 \end{eqnarray*}
$H_{atoms}=\frac{\hbar\Omega}{2}(x_{A}^{2}+p_{A}^{2})\ $ refers to
Zeeman-splitting of the atomic ground state causing Larmor
precession of the transverse spin components represented by
$x_{A}$ and $p_{A}$.  The interaction terms $V_{1}$ and $V_{2}$
account for the off-resonant scattering interaction in the first
and second passage of the pulse respectively. They are given by
\begin{eqnarray*}
V_{1}=\frac{\hbar\kappa}{\sqrt{T}}p_{A}p_{L}(0)\ \ \textrm{and}\ \
V_{2}=\frac{\hbar\kappa}{\sqrt{T}}x_{A}x_{L}(d)\ ,
\end{eqnarray*}
where $T$ is the duration of the pulse. $V_{1}$ was already
introduced. $V_{2}$ basically describes the same kind of
interaction, but due to the changed geometry in the second pass
atomic quadratures are interchanged $p_{A} \rightarrow x_{A}$.
Since the beam is sent through a quarter wave plate between it's
passes through the atomic sample, light quadratures are
interchanged as well $p_{L} \rightarrow x_{L}$. The arguments of
the light-operators in $V_{1}$ and $V_{2}$ indicate that the first
scattering interaction occurs at $r=0$, while the second
interaction happens after the light has travelled some distance
$d$ in the small loop between the mirrors. The length of the laser
pulse is hereby supposed to be large compared with the distance
within the loop. In typical experiments pulses of a length of
several hundred km are used, therefore the pulse encounters itself
in the sample \footnote{Due to thermal motion each atom enters and
leaves the interaction volume covered by both beams several times,
such that the beams couple effectively both to the atomic center
of mass mode with an averaged constant coupling strength}.
$H_{light}$ represents free propagation of light. It acts on light
quadratures like
$\partial_{t}x_{L}(r)=\frac{i}{\hbar}[H_{light},x_{L}(r)]\cong-c\partial_{r}x_{L}(r)$,
which is a suitable approximation for the light modes defined in
(\ref{Light modes}).
Evaluating the Heisenberg equations gives
\begin{eqnarray*}
  \partial_{t}x_{A}(t)&=&\Omega
  p_{A}(t)+\frac{\kappa}{\sqrt{T}}p_{L}(0,t)\ ,\\
  \partial_{t}p_{A}(t)&=& -\Omega
  x_{A}(t)-\frac{\kappa}{\sqrt{T}}x_{L}(d,t)\ ,\\
  (\partial_{t}+c\partial_{r})x_{L}(r,t)&=& \frac{\kappa c}{\sqrt{T}}p_{A}(t)\delta(r)\ ,\\
  (\partial_{t}+c\partial_{r})p_{L}(r,t)&=&- \frac{\kappa
  c}{\sqrt{T}}x_{A}(t)\delta(r-d)\ .
\end{eqnarray*}
By performing the variable transformation $\xi=ct-r$ we obtain the
Maxwell-Bloch equations
\begin{eqnarray}
  \partial_{t}x_{A}(t)&=&\Omega
    p_{A}(t)+\frac{\kappa}{\sqrt{T}}\overline{p}_{L}(ct,t),\label{bloch equation x memory}\\
  \partial_{t}p_{A}(t)&=&-\Omega
    x_{A}(t)-\frac{\kappa}{\sqrt{T}}\overline{x}_{L}(ct-d,t), \label{bloch equation p memory}\\
  \partial_{t}\overline{x}_{L}(\xi,t)&=&\frac{\kappa
    c}{\sqrt{T}}p_{A}(t)\delta(ct-\xi), \label{maxwell equation x memory}\\
  \partial_{t}\overline{p}_{L}(\xi,t)&=&-\frac{\kappa
    c}{\sqrt{T}}x_{A}(t)\delta(ct-\xi-d).\label{maxwell equation p memory}
\end{eqnarray}
Light modes in new variables are denoted by a bar
$\overline{x}_{L}(\xi,t)=x_{L}(ct-\xi,t)$. The light variable
argument $\xi$ refers to a coordinate system which is fixed on the
light pulse. It allows one to denote easily particular pieces on the
pulse. At a certain instant of time $\xi$ labels the pieces
according to their position starting with the piece, which enters
the atomic sample first.
\\
This set of coupled differential equations has now to be solved.
As a first step we treat the  equations for light.
In the first pass a $p_{L}p_{A}$ -interaction occurs and $x_{L}$
picks up some $p_{A}$ contribution. The delta function in
(\ref{maxwell equation x memory}) reflects the fact that a certain
piece $\xi$ of the pulse gets a contribution from the atomic state
at $t=\xi/c$ (which is the instant of time the piece in
consideration passes by). In the second pass a $x_{L}x_{A}$
-interaction occurs, and the atomic $x$ -quadrature is written onto
$p_{L}$. A piece $\xi$ of the pulse, which interacted with $p_{L}$
at time $\xi/c$ gets a contribution from $x_{L}$ after it has
traveled a distance $d$ in the loop. Therefore the atomic $x$
quadrature is picked up at $t=\xi/c+d/c$ which is indicated by the
delta-function in equation (\ref{maxwell equation p memory}). By
integrating equations (\ref{maxwell equation x memory}) and
(\ref{maxwell equation p memory}) formally these delta functions
turn into Heaviside functions,
\begin{eqnarray*}
  \overline{x}_{L}(\xi,t)\!\!\!&=&\!\! \overline{x}_{L}(\xi,0)+\!\frac{\kappa c}{\sqrt{T}}\int_{0}^{t}d\tau p_{A}(\tau)\delta(c\tau-\xi)\\
  \!\!\!&=&\!\!\overline{x}_{L}(\xi,0)+\!\frac{\kappa}{\sqrt{T}}p_{A}(\xi /c)\Theta(t-\xi/c),\\
  \overline{p}_{L}(\xi,t)\!\!\!&=&\!\!\overline{p}_{L}(\xi,0)-\!\frac{\kappa c}{\sqrt{T}}\int_{0}^{t}d\tau
  x_{A}(\tau)\delta(c\tau-\xi-d)\\
  \!\!&=&\!\!\overline{p}_{L}(\xi,0)-\!\frac{\kappa}{\sqrt{T}}x_{A}(\xi/c+d /c)\Theta(t-\xi/c-d /c).\\
\end{eqnarray*}
Now $\overline{p}_{L}(ct,t)$ and $\overline{x}_{L}(ct-d,t)$ are
calculated, since these expressions have to be substituted into the
atomic differential equations (\ref{bloch equation x memory}) and
(\ref{bloch equation p memory}). The fact that the arguments are
different for $x_{L}$ and $p_{L}$ can be understood by considering
the processes going on in the course of the double pass scheme.
During the $p_{L}p_{A}$-interaction in the first passage $x_{A}$
picks up some $p_{L}$ contribution. If we consider this process at a
certain instant of time $t$, the relevant piece of the pulse is the
one passing $r=0$. It is denoted by $\xi=ct-r=ct$. $p_{A}$ is acted
upon by $x_{L}$ in the second pass by the piece of the pulse which
passes $r=d$ at time $t$. So it gets a contribution from
$\overline{p}_{L}(ct-d,t)$. One finds
\begin{eqnarray*}
  \overline{x}_{L}(ct-d,t)&=&\overline{x}_{L}(ct-d,0)+\frac{\kappa}{\sqrt{T}}p_{A}(t-d/c)\Theta(d/c)\\
  &=&\overline{x}_{L}(ct-d,0)+\frac{\kappa}{\sqrt{T}}p_{A}(t-d/c),\\
  \overline{p}_{L}(ct,t)&=&
  \overline{p}_{L}(ct,0)-\frac{\kappa}{\sqrt{T}}x_{A}(t+d/c)\Theta(-d/c)\\
  &=& \overline{p}_{L}(ct,0).
\end{eqnarray*}
Note that $\overline{p}_{L}(ct,t)$ is conserved. This feature is
due to the time-delay in the loop and will turn out to be crucial
for the characteristic exponential behavior of the whole scheme.
After inserting these results into (\ref{bloch equation x memory})
and (\ref{bloch equation p memory}) the atomic differential
equations read
\begin{eqnarray*}
  \partial_{t}x_{A}(t)\!\!&=&\!\! \Omega p_{A}(t)
    \!+\!\frac{\kappa}{\sqrt{T}}\overline{p}_{L}(ct,0),\\
  \partial_{t}p_{A}(t)\!\!&=&\!\! -\Omega x_{A}(t)
   \!-\!\frac{\kappa}{\sqrt{T}}\overline{x}_{L}(ct-d,0)\!-\!\frac{\kappa^{2}}{T}p_{A}(t-d/c).
\end{eqnarray*}
Now we assume $d/c \ll \Omega^{-1}$, such that the elapsed time
during the run in the loop is definitely much shorter than any
other relevant process. $d/c$ can be assumed to be of the order of
$ns$ while atoms rotate slowly with a Larmor period of the order
of $\mu s$. With this approximation
\begin{eqnarray}\label{atomic differential equation memory}
\partial_{t}\!\left(%
\begin{array}{c}
  \! x_{A}(t)\!\\
 \! p_{A}(t)\! \\
\end{array}%
\right)\!\!&=&\!\! \left\{ \Omega \left(%
\begin{array}{cc}
  \!0 & 1\! \\
  \!-1 & 0\!\\
\end{array}%
\right) -\frac{\kappa^{2}}{T}\left(%
\begin{array}{cc}
  \!0 & 0\! \\
 \!0 & 1\! \\
\end{array}%
\right)\right\}
\left(%
\begin{array}{c}
 \!x_{A}(t)\! \\
  \!p_{A}(t)\! \\
\end{array}%
\right)\nonumber\\
\!\!&&\!\!+\frac{\kappa}{\sqrt{T}} \left(%
\begin{array}{c}
  \!\overline{p}_{L}(ct,0)\! \\
 \! -\overline{x}_{L}(ct,0)\! \\
\end{array}%
\right).
\end{eqnarray}
This differential equation consists of a homogeneous part and a
driving term. The first term of the homogeneous part - being
proportional to the Larmor frequency - reflects the fact that
atoms turn with $\Omega$ in the external magnet field. The second
term in the homogeneous part represents damping of $p_{A}$.
Although only one quadrature is damped, the effect is distributed
among both quadratures by Larmor precession. This leads to a
symmetry between $x$ and $p$, which is a characteristic feature of
our proposal. The solution to the differential equation is
\begin{eqnarray*}
 \left(%
\begin{array}{c}
   x_{A}(t)\\
  p_{A}(t) \\
\end{array}%
\right)
&=&A(t)\left(%
\begin{array}{c}
  x_{A}(0) \\
  p_{A}(0) \\
\end{array}%
\right)\\
&&+A(t) \frac{\kappa}{\sqrt{T}}\int_{0}^{t}d\tau\
A^{-1}(\tau)\left(%
\begin{array}{c}
   \overline{p}_{L}(c \tau,0) \\
   -\overline{x}_{L}(c \tau,0) \\
\end{array}%
\right), \\
\end{eqnarray*}
where $A(t)=e^{G t}$ \ ,\  $G={\Omega \left(%
\begin{array}{cc}
  0 & 1 \\
  -1 & 0 \\
\end{array}%
\right) -\frac{\kappa^{2}}{T}\left(%
\begin{array}{cc}
  0 & 0 \\
  0 & 1 \\
\end{array}%
\right)\ } $\\
\\
is the homogeneous solution. We suppose $\Omega T\gg \kappa^{2}$,
which matches experimental conditions, since typically $\Omega T
\approx 300$ while $\kappa^{2}$ is of order unity. With this
assumption
\begin{eqnarray*}
A(t)=e^{-\frac{\kappa^{2}t}{2 T}} R^{-1}(t)\ ,
\end{eqnarray*}
where $R^{-1}(t)$ is an orthogonal matrix,
\begin{eqnarray*}
R^{-1}(t)=\left(%
\begin{array}{cc}
  \cos(\Omega t) & \sin(\Omega t) \\
  -\sin(\Omega t) & \cos(\Omega t)\\
\end{array}%
\right).
\end{eqnarray*}
The inverse is taken for later convenience. Therefore the atomic time evolution is given by \\
\begin{eqnarray}\label{atomic time evolution memory}
  \left(%
\begin{array}{c}
  \!\! x_{A}(t)\!\!\\
   \!\!p_{A}(t) \!\!\\
\end{array}%
\right)\!\!\!\!&=&\!\!e^{-\frac{\kappa^{2}t}{2 T}}R^{-1}(t)\left(%
\begin{array}{c}
   x_{A}^{in} \\
   p_{A}^{in} \\
\end{array}%
\right) \\
 \!\!&&\!\!+ e^{-\frac{\kappa^{2}t}{2 T}}R^{-1}(t)\!
\frac{\kappa}{\sqrt{T}}\!
 \int_{0}^{t}\!\!\!d\tau e^{\frac{\kappa^{2}\tau}{2 T}}\!
R(\tau)\! \left(%
\begin{array}{c}
  \!\!\overline{p}_{L}(c \tau,0) \!\! \\
  \!\!-\overline{x}_{L}(c \tau,0) \!\! \\
\end{array}%
\right)\! .\nonumber
\end{eqnarray}
Now the atomic output quadratures $x^{out}_{A}=x_{A}(T)$ and
$p^{out}_{A}=p_{A}(T)$ can be directly written down. With the
assumption $\Omega T=2 \pi n$ for some natural number $n$, \\
\begin{eqnarray*}
\left(%
\begin{array}{c}
   x^{out}_{A} \\
   p^{out}_{A} \\
\end{array}%
\right)&=&e^{-\frac{\kappa^{2}}{2}}\left(%
\begin{array}{c}
   x^{in}_{A} \\
   p^{in}_{A} \\
\end{array}%
\right)\\
&&+ e^{-\frac{\kappa^{2}}{2}} \frac{\kappa}{\sqrt{T}}
\int_{0}^{T}dt\  e^{\frac{\kappa^{2}t}{2
T}}R(t) \left(%
\begin{array}{c}
   \overline{p}_{L}(c t,0)  \\
  -\overline{x}_{L}(c t,0)  \\
\end{array}%
\right).
\end{eqnarray*}
The atomic output-quadratures consist of some atomic input
contribution which is damped exponentially with $\kappa^{2}$ and
an additional light contribution which they pick up during the
scattering interaction. The definition of the appropriate
light-mode can be taken from this result right away,\\
\begin{equation}\label{plus mode memory}
   \left(%
\begin{array}{c}
 \! x^{in}_{L+}\! \\
 \! p^{in}_{L+}\! \\
\end{array}%
\right)=\frac{\kappa}{\sqrt{T}\sqrt{e^{\kappa^{2}}-1}}\int_{0}^{T}dt\
e^{\frac{\kappa^{2}t}{2 T}}R(t) \left(%
\begin{array}{c}
  \overline{p}_{L}(c t,0)  \\
  -\overline{x}_{L}(c t,0)  \\
\end{array}%
\right),
\end{equation}\\
where the prefactor assures normalization such that
$[x^{in}_{L+},p^{in}_{L+}]=i$. This new defined light mode is
essentially the upper sideband mode $x_{u s}$, $p_{u s}$, which is
given by
\begin{eqnarray}\label{upper sideband}
  \left(%
    \begin{array}{c}
      x_{u s}^{in}  \\
      p_{u s}^{in}  \\
    \end{array}%
  \right)
 &=& \frac{1}{\sqrt{T}}\int_{0}^{T}dt\ R(t)\left(%
  \begin{array}{c}
    \overline{p}_{L}(ct,0)  \\
    -\overline{x}_{L}(ct,0)  \\
 \end{array}%
 \right).
\end{eqnarray}
The only difference is given by the fact that the stored mode
$x^{in}_{L+},p^{in}_{L+}$ is defined with a slowly varying
envelope of the form $\exp(+\kappa^2t/2T)$. The index $"+"$ refers
to the sign of the argument in this exponential function (later on
we will also have to deal with corresponding $"-"$ modes). With
use of (\ref{plus mode memory}) the atomic input-output relations
can be written in a compact form
\begin{equation}\label{atomic input-output relations memory}
\left(%
\begin{array}{c}
  x^{out}_{A} \\
  p^{out}_{A} \\
\end{array}%
\right)=e^{-\frac{\kappa^{2}}{2}}\left(%
\begin{array}{c}
  x^{in}_{A} \\
  p^{in}_{A} \\
\end{array}%
\right)+ \sqrt{1-e^{-\kappa^{2}}}\left(%
\begin{array}{c}
  x_{L +}^{in}  \\
  p_{L +}^{in}  \\
\end{array}%
\right).
\end{equation}
These equations describe the write-in process for a signal, which
is encoded at the mode described above. Remarkably, mapping of
such a quantum state of light onto atoms approaches perfect
read-in exponentially in the coupling strength. This arises from
the fact, that in the course of the double pass scattering
interaction $x_{L}$ picks up some contribution from the atomic $p$
-quadrature, while $p_{L}$ in contrast is conserved. Therefore we
do not get a rotating term in the basic differential equation
(\ref{atomic differential equation memory}), which would lead to
sines and cosines in the solution, as we would expect for a beam
splitter like interaction, but an exponential effect, which is
characteristic for the setup.
\subsection{Read-out}\label{Read-out}
In order to perform the read-out, a pulse of light has to be sent
through the double-pass setup, just like for the write-in
procedure, but since we are now looking at the reverse process,
the appropriate light mode for this task has to be accented in an
inverse fashion. While in the write-in process the rear part of
the pulse was emphasized, now the front part of the pulse has to
be weighted in order to pick up atomic information best. As the
exponent in the mode definition is negative this read-out mode
will be denoted by a minus sign. Since we now deal with a new beam
of light which is independent from the write-in pulse, read-out
beam variables carry an accent,
\begin{equation*}
     \left(%
\begin{array}{c}
 \! \acute{x}^{in}_{L-}\! \\
 \! \acute{p}^{in}_{L-}\!  \\
\end{array}%
\right)\!=\!\frac{\kappa}{\sqrt{T}\sqrt{1-e^{-\kappa^{2}}}}\int_{0}^{T}\!\!dt\
e^{-\frac{\kappa^{2}t}{2 T}}R(t) \left(%
\begin{array}{c}
  \!\acute{\overline{p}}_{L}(c t,0)\!  \\
 \! -\acute{\overline{x}}_{L}(c t,0)\!  \\
\end{array}%
\right)\!,
\end{equation*}
with a new normalization constant $\kappa/
(\sqrt{T}\sqrt{1-e^{-\kappa^{2}}})$.
The input-output relations for this mode can be derived by changing
the time argument of light operators from $0$ to $T$, reflecting the
fact that we now look at the light quadratures after the whole pulse
run through the atomic sample
\begin{equation*}
    \left(%
\begin{array}{c}
  \!\acute{x}^{out}_{L-}\! \\
  \!\acute{p}^{out}_{L-}\!  \\
\end{array}%
\right)\!=\!\frac{\kappa}{\sqrt{T}\sqrt{1-e^{-\kappa^{2}}}}\int_{0}^{T}\!\!dt e^{-\frac{\kappa^{2}t}{2 T}}R(t) \left(%
\begin{array}{c}
 \! \acute{\overline{p}}_{L}(c t,T) \! \\
 \! -\acute{\overline{x}}_{L}(c t,T)\!  \\
\end{array}%
\right)\!.
\end{equation*}
\\To evaluate this expression in terms of input-operators the integrated versions of equations (\ref{maxwell equation x memory}) and
(\ref{maxwell equation p memory})
\begin{eqnarray*}
    \acute{\overline{x}}_{L}(\xi,t)&=&\acute{\overline{x}}_{L}(\xi,0)+\frac{\kappa}{\sqrt{T}}p_{A}(\xi /c)\Theta (t-\xi /c),\\
    \acute{\overline{p}}_{L}(\xi,t)&=&\acute{\overline{p}}_{L}(\xi,0)-\frac{\kappa}{\sqrt{T}}x_{A}(\xi
    /c)\Theta(t-\xi/c).
\end{eqnarray*}
 are used. Therefore
\begin{eqnarray*}
    \left(%
\begin{array}{c}
  \!\acute{x}^{out}_{L-} \!\\
  \!\acute{p}^{out}_{L-} \! \\
\end{array}%
\right)\!\!\!&=&\!\!\!\frac{\kappa}{\sqrt{T}\sqrt{1-e^{-\kappa^{2}}}}\int_{0}^{T}\!\!\!dt\ e^{-\frac{\kappa^{2}t}{2 T}}R(t)\\
\!\!&&\!\!\left[ \left(%
\begin{array}{c}
  \acute{\overline{p}}_{L}(ct,0)    \\
  -\acute{\overline{x}}_{L}(ct,0)  \\
\end{array}%
\right) -\frac{\kappa}{\sqrt{T}}
\left(%
\begin{array}{c}
  \!x_{A}(t)\!\\
  \!p_{A}(t)\!\\
\end{array}%
\right)\!\! \Theta(T-t)
\right],\\
\!\!&=&\!\!\!\!\left(%
\begin{array}{c}
  \!\acute{x}^{in}_{L-}\! \!\\
 \! \acute{p}^{in}_{L-}\! \! \\
\end{array}%
\right)\!-\frac{\kappa^{2}}{T\sqrt{1-e^{-\kappa^{2}}}}\int_{0}^{T}\!\!\!\!dt
e^{-\frac{\kappa^{2}t}{2 T}}R(t)\!
\left(%
\begin{array}{c}
  \!\!x_{A}(t) \!\! \\
  \!\!p_{A}(t) \!\! \\
\end{array}%
\right)\!.
\end{eqnarray*}
Now the atomic time evolution (\ref{atomic time evolution memory})
has to be inserted. The resulting expression can be simplified by
interchanging the order of the double-integral
$\int^{T}_{0}dt\int^{t}_{0}d\tau\rightarrow\int^{T}_{0}d\tau\int^{T}_{\tau}dt$.
With help of equation (\ref{plus mode memory}) the read-out output
can then be written as a sum of an atomic contribution and some
contribution from the plus-mode.
\begin{equation}\label{light input-output relations memory}
 \left(%
  \begin{array}{c}
   \acute{x}^{out}_{L-} \\
   \acute{p}^{out}_{L-}  \\
   \end{array}%
 \right)=-\sqrt{1-e^{-\kappa^{2}}}
  \left(%
  \begin{array}{c}
   x^{in}_{A} \\
   p^{in}_{A}  \\
   \end{array}%
 \right)
 +e^{-\frac{\kappa^{2}}{2}}
 \left(%
  \begin{array}{c}
   \acute{x}^{in}_{L+} \\
   \acute{p}^{in}_{L+}  \\
   \end{array}%
 \right)
\end{equation}
Note that this expression resembles the formula for the write-in
procedure with the roles of light- and atomic modes interchanged.
\subsection{Fidelity for the complete state transfer}\label{Fidelity for the complete state transfer}
The fidelity for the complete state transfer is given by the
overlap of the initial input state and the final output state
after storage and subsequent retrieval. By inserting the output of
the write-in procedure (\ref{atomic input-output relations
memory}) into the read-out equation (\ref{light input-output
relations memory}) one obtains
\begin{eqnarray}\label{complete state transfer}
    x^{fin}_{L-}\!\!&=&\!\!-(1-e^{-\kappa^{2}}\!\!)x^{in}_{L+}-e^{-\frac{\kappa^{2}}{2}}
    \sqrt{1-e^{-\kappa^{2}}}x^{in}_{A}
     +e^{-\frac{\kappa^{2}}{2}}\acute{x}^{in}_{L-},\nonumber\\
     p^{fin}_{L-}\!\!&=&\!\!-(1-e^{-\kappa^{2}}\!\!)p^{in}_{L+}-e^{-\frac{\kappa^{2}}{2}}
     \sqrt{1-e^{-\kappa^{2}}}p^{in}_{A}
    + e^{-\frac{\kappa^{2}}{2}}\acute{p}^{in}_{L-}.\nonumber\\
\end{eqnarray}
For infinite coupling $\kappa^{2}$ the original input-signal is
retrieved within the final quadratures of the read-out pulse
$x_{L-}^{fin}=-x_{L+}^{in}$ and $p_{L-}^{fin}=-p_{L+}^{in}$, while
the noise terms (atomic and
read-out beam input contributions) vanish.\\
The quantum state to be stored is supposed to be unknown. It is
assumed to be taken from a certain set of possible input-states.
In the following two subsections we will consider coherent input
states and light qubits respectively. We will first calculate the
fidelity for a single state transfer and take the average over the
complete set of possible input states in the next step in each
case. The results will be compared to the corresponding classical
limits, i.e. the maximum average fidelity, that can be achieved by
classical means \cite{Quantum benchmark 1, Quantum benchmark 2,
Qubit fidelity 1, Qubit fidelity 2, Qubit fidelity 3, Qubit
fidelity 4}.
\subsubsection{Fidelity for coherent input states}\label{Fidelity for coherent input states}
We first consider storage of a coherent state of light. The
overlap between an initial state with quadratures $x_{L +}^{in}$,
$p_{L +}^{in}$ and the final state with $\acute{x}_{L-}^{fin}$,
$\acute{p}_{L-}^{fin}$ is given by
\begin{eqnarray}\label{coherent fidelity}
F_{coh}&=&\frac{2}{\sqrt{[1+2(\Delta
\acute{x}^{fin}_{L-})^{2}][1+2(\Delta \acute{p}^{fin}_{L-})^{2}]}}\nonumber\\
&&e^{-\frac{(\langle x_{L +}^{in}\rangle+\langle
\acute{x}_{L-}^{fin}\rangle)^{2}}{1+2(\Delta
\acute{x}^{fin}_{L-})^{2}}-\frac{(\langle p_{L
+}^{in}\rangle+\langle
\acute{p}_{L-}^{fin}\rangle)^{2}}{1+2(\Delta
\acute{p}^{fin}_{L-})^{2}}}.
\end{eqnarray}
The expectation values and variances of the final light state
follow directly from (\ref{complete state transfer}). Since the
atoms and the read-out plus mode are initially in a vacuum state
we have $ \langle x^{fin}_{L-}\rangle=-(1-e^{-\kappa^{2}})\langle
x^{in}_{L+}\rangle$ and $\langle
p^{fin}_{L-}\rangle=-(1-e^{-\kappa^{2}})\langle
p^{in}_{L+}\rangle$, while the variances are given by $ (\Delta
x^{fin}_{L-})^{2}= (\Delta p^{fin}_{L-})^{2}=\frac{1}{2}$, as one
expects for a passive transformation. Therefore
\begin{eqnarray*}
F_{coh}=e^{-\frac{1}{2}(\langle x^{in}_{L+}\rangle^{2}+\langle
p^{in}_{L+}\rangle^{2})e^{-2\kappa^{2}}}\ .
\end{eqnarray*}
Now the average fidelity is computed by averaging over the
complete set of all possible coherent input states. For this
purpose the amplitudes $x_{L +}^{in}$ and $p_{L +}^{in}$ are
assumed to be taken according to a Gaussian distribution centered
at zero with a certain width $n$.
\begin{eqnarray*}
\overline{F}_{coh}(n,\kappa)&=&\frac{1}{2 \pi n}\ \int \! \int
d\langle x^{in}_{L +}\rangle d\langle p^{in}_{L +}\rangle\
e^{-\frac{\langle x^{in}_{L +}\rangle^{2}+\langle p^{in}_{L
+}\rangle^{2}}{2 n}}\\
&&F_{coh}(\langle x^{in}_{L +}\rangle , \langle p^{in}_{L
+}\rangle ,\kappa)\ ,\\
&=&\frac{1}{1+e^{-2\kappa^{2}}n} .\\
\end{eqnarray*}
Figure \ref{Fidelities}a shows the average fidelity for different
widths corresponding to mean photon numbers of the distribution.
The corresponding classical limit $\bar{F}^{cl}_{coh}=\frac{2n
+1}{4n+1}$ \cite{Quantum benchmark 1, Quantum benchmark 2} is
marked by a cross on each curve.
\subsubsection{Fidelity for light qubits}\label{Fidelity for light qubits}
\begin{figure}[btp]
\begin{center}
\includegraphics[width=6cm]{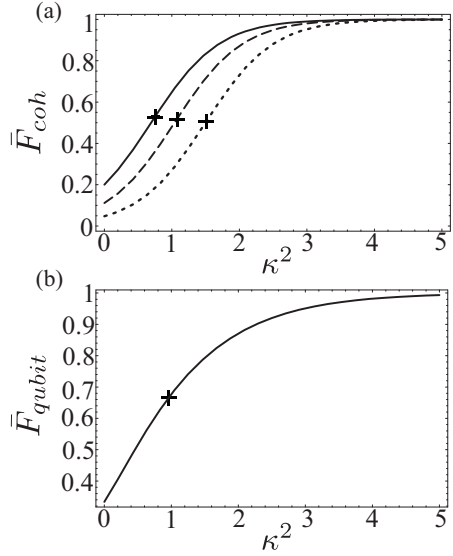}
\caption{Average fidelity for write-in and subsequent read-out of
a light state versus coupling $\kappa^{2}$. Crosses indicate the
classical limit in each case. \textbf{(a)} Average fidelity for
coherent light states according to distributions with different
mean photon numbers (solid line: n=4, dashed line: n=8, dotted
line: n=20) \textbf{(b)} Average fidelity for light qubits.
\label{Fidelities}}
\end{center}
\end{figure}

Now the fidelity for light-qubits is calculated. The light-qubit
input state is represented by
\begin{equation*}
    |\Psi^{in}\rangle=(\alpha+\beta
    a_{L+}^{\dag \ in})|vac\rangle,
\end{equation*}
where $a_{L+}^{\dag \ in}=\frac{1}{\sqrt{2}}(x^{in}_{L+}-i
p^{in}_{L+})$ is the creation operator for a photon in the
write-in mode.
The write-in and read-out procedure is given by a passive
transformation $U$
\begin{eqnarray}
  \nonumber |\Psi^{fin}\rangle=U  |\Psi^{in}\rangle&=&(\alpha+\beta
    Ua_{L+}^{\dag \ in})|vac\rangle\\ \nonumber
    &=&(\alpha+\beta
    Ua_{L+}^{\dag \ in}U^{\dag})|vac\rangle\\
    &=&(\alpha+\beta
    a_{L-}^{\dag\ fin})|vac\rangle \ ,  \label{final qubit state}
\end{eqnarray}
where $U|vac\rangle=|vac\rangle$ was used. Here $a_{L-}^{\dag \
fin}=\frac{1}{\sqrt{2}}(x^{fin}_{L-}-i p^{fin}_{L-})$ is the
creation operator after mapping and subsequent retrieval. It can be
directly calculated, since the complete input-output relations for
the light-quadratures are known. With use of equations
(\ref{complete state transfer}) one finds
\begin{equation}
a^{\dag \ fin}_{L-}\!\!=\!\!-(1-e^{-\kappa^{2}}\!)a^{\dag \
in}_{L+}-e^{-\frac{\kappa^{2}}{2}}
   \sqrt{1-e^{-\kappa^{2}}}a^{\dag \ in}_{A}
    +e^{-\frac{\kappa^{2}}{2}}\acute{a}^{\dag \
    in}_{L-},\label{complete state transfer for creation operators}
\end{equation}
where $a^{\dag \ in}_{L+}$, $a^{\dag \ in}_{A}$ and $\acute{a}^{\dag
\ in}_{L-}$ refer to the light state to be stored, the atoms and the
read-out mode respectively. The fidelity is given by the state
overlap between $|\Psi^{fin}\rangle$ and the optimal final state
$|\Psi^{fin}_{opt}\rangle=(\alpha-\beta a_{L+}^{\dag \
in})|vac\rangle$. By inserting (\ref{complete state transfer for
creation operators}) into expression (\ref{final qubit state})
$F_{qubit}$ can easily be determined. One obtains
\begin{eqnarray*}
F_{qubit}=|\langle\Psi^{fin}|\Psi^{fin}_{opt}\rangle|^{2}
=|(|\alpha|^{2}+\{1-e^{-\kappa^{2}}\}|\beta|^{2})|^{2}.
\end{eqnarray*}
The average fidelity is calculated by setting
$\alpha=\cos(\frac{\theta}{2})$ and
$\beta=\sin(\frac{\theta}{2})e^{i \phi}$ and integrating over the
whole Bloch-sphere,
\begin{eqnarray*}
\overline{F}_{qubit}(\kappa)&=&\frac{1}{4 \pi}\int^{\pi}_{0}d
\theta \int^{2 \pi}_{0}d \phi
\sin(\theta)F_{qubit}(\theta,\phi)\\
&=&1-e^{- \kappa^{2}}+\frac{1}{3}e^{-2 \kappa^{2}}.
\end{eqnarray*}
Figure \ref{Fidelities}b shows this result. The maximal average
fidelity that can be achieved for qubit states by a classical
strategy $\bar{F}^{cl}_{qubit}=\frac{2}{3}$ \cite{Qubit fidelity
1, Qubit fidelity 2, Qubit fidelity 3} is indicated by a cross.
\section{Exponential two mode squeezing}\label{Two mode squeezing}
The interaction which governs the squeezing scheme pictured in
figure \ref{Setups}(b), is given by
\begin{eqnarray*}
 \tilde{H}=H_{atoms}+H_{light}+V_{1}-V_{2}\ .
 \end{eqnarray*}
This Hamiltonian differs from the one used in the memory section
just by a sign in the interaction term referring to the second
passage. The pulse runs along $-\hat{y}$ in the second pass of the
squeezing scheme (instead of $\hat{y}$ in the previous case) and
sees therefore $-x_{A}$. Hence we have the minus sign in front of
$V_{2}$ for the new setup.
\subsection{Input-output relations}
 The atomic input-output relations can
now be derived in complete analogy to section \ref{Write-in}. By
evaluating the Heisenberg equations as above we get
\begin{eqnarray*}
\partial_{t}\left(%
\begin{array}{c}
   \!x_{A}(t)\!\\
  \!p_{A}(t)\! \\
\end{array}%
\right)&=& \left\{ \Omega \left(%
\begin{array}{cc}
 \! 0 & 1 \!\\
  \!-1 & 0\! \\
\end{array}%
\right) +\frac{\kappa^{2}}{T}\left(%
\begin{array}{cc}
 \! 0 & 0 \!\\
  \!0 & 1 \!\\
\end{array}%
\right)\right\}
\left(%
\begin{array}{c}
  \!x_{A}(t)\! \\
 \! p_{A}(t)\! \\
\end{array}%
\right)\\
&&+\frac{\kappa}{\sqrt{T}} \left(%
\begin{array}{c}
  \!\overline{p}_{L}(ct,0)\! \\
 \! \overline{x}_{L}(ct,0)\! \\
\end{array}%
\right).
\end{eqnarray*}
With the usual approximation $\kappa^{2} \ll 2 \Omega T$ we obtain
\begin{eqnarray*}
  \left(%
\begin{array}{c}
  \!\! x_{A}(t)\!\!\\
  \!\! p_{A}(t)\!\! \\
\end{array}%
\right)\!\!&=&\!\!e^{\frac{\kappa^{2}t}{2 T}}R^{-1}(t)\left(%
\begin{array}{c}
  \!\! x_{A}^{in}\!\! \\
   \!\!p_{A}^{in}\!\! \\
\end{array}%
\right)\\
\!\!&& \!\!+e^{\frac{\kappa^{2}t}{2 T}}R^{-1}(t)
\frac{\kappa}{\sqrt{T}}\!\! \int_{0}^{t}\!\!\!d\tau
e^{-\frac{\kappa^{2}\tau}{2 T}}
R(\tau) \!\!\left(%
\begin{array}{c}
 \!\! \overline{p}_{L}(c \tau,0) \!\! \\
 \!\! \overline{x}_{L}(c \tau,0) \!\! \\
\end{array}%
\right).
\end{eqnarray*}
These equations are in a significant way different from the atomic
time evolution (\ref{atomic time evolution memory}) in the memory
scheme. Note first the signs in the arguments of the exponential
functions. We now have exponential enhancement of the atomic input
instead of exponential damping. Furthermore light is involved in
form of a minus mode in the atomic input-output relations because of
the minus sign in the exponent within the integral. Note second,
that the minus sign, which was present in front of
$\overline{x}_{L}(ct,0)$ in the memory scheme, does not appear in
this case. Therefore the lower sideband
\begin{eqnarray}\label{lower sideband}
  \left(%
    \begin{array}{c}
      p_{l s}^{in}  \\
      x_{l s}^{in}  \\
    \end{array}%
  \right)
 &=& \frac{1}{\sqrt{T}}\int_{0}^{T}dt\ R(t)\left(%
  \begin{array}{c}
    \overline{p}_{L}(ct,0)  \\
    \overline{x}_{L}(ct,0)  \\
 \end{array}%
 \right)
\end{eqnarray}
is involved instead of the upper one (\ref{upper sideband}). Hence
the minus mode showing up in the atomic time evolution is defined
slightly differently from the memory section
\begin{eqnarray*}
    \left(%
\begin{array}{c}
 \!\! \tilde{p}^{in}_{L-} \!\!\\
  \!\!\tilde{x}^{in}_{L-} \!\! \\
\end{array}%
\right)\!\!=\!\!\frac{\kappa}{\sqrt{T} \sqrt{1-e^{-\kappa^{2}}}}
\int_{0}^{T}\!\!dte^{-\frac{\kappa^{2}t}{2 T}}R(t) \left(%
\begin{array}{c}
  \!\!\overline{p}_{L}(c t,0)\!\!  \\
  \!\!\overline{x}_{L}(c t,0)\!\!  \\
\end{array}%
\right).
\end{eqnarray*}
With use of this definition and the assumption $\Omega T =2 \pi n
$ for some natural number $n$, the atomic input-output relations
read
\begin{eqnarray}\label{atomic input-output relations squeezer}
    \left(%
\begin{array}{c}
  x^{out}_{A} \\
  p^{out}_{A} \\
\end{array}%
\right)=e^{\frac{\kappa^{2}}{2}}
\left(%
\begin{array}{c}
  x^{in}_{A} \\
  p^{in}_{A} \\
\end{array}%
\right)+\sqrt{e^{\kappa^{2}}-1}
\left(%
\begin{array}{c}
  \tilde{p}^{in}_{L-} \\
  \tilde{x}^{in}_{L-}\\
\end{array}%
\right).
\end{eqnarray}
Light input-output relations for this process can be derived in
analogy to the procedure in section \ref{Read-out}. The inverse
accented counter-part of the light mode used in the atomic
evolution is given by
\begin{eqnarray*}
    \left(%
\begin{array}{c}
  \tilde{p}^{in}_{L+} \\
  \tilde{x}^{in}_{L+}  \\
\end{array}%
\right)=\frac{\kappa}{\sqrt{T} \sqrt{e^{\kappa^{2}}-1}}
\int_{0}^{T}dt\
e^{\frac{\kappa^{2}t}{2 T}}R(t) \left(%
\begin{array}{c}
  \overline{p}_{L}(c t,0)  \\
  \overline{x}_{L}(c t,0)  \\
\end{array}%
\right).
\end{eqnarray*}
We refer now to the same pulse as in (\ref{atomic input-output
relations squeezer}), while in the derivation of the input-output
relations for atoms and light in section \ref{Quantum memory} two
independent beams were considered. We obtain
\begin{eqnarray}\label{light input-output relations squeezer}
  \left(%
\begin{array}{c}
  \tilde{p}^{out}_{L+} \\
  \tilde{x}^{out}_{L+}  \\
\end{array}%
\right)=\sqrt{e^{\kappa^{2}}-1}\left(%
\begin{array}{c}
  x_{A}^{in} \\
  p_{A}^{in} \\
\end{array}%
\right)+ e^{\frac {\kappa^{2}}{2}}
\left(%
\begin{array}{c}
  \tilde{p}^{in}_{L-} \\
  \tilde{x}^{in}_{L-}  \\
\end{array}%
\right).
\end{eqnarray}
Please note that the input-output relations (\ref{atomic
input-output relations squeezer}) and (\ref{light input-output
relations squeezer}) are active versions of (\ref{atomic
input-output relations memory}) and (\ref{light input-output
relations memory}) respectively.
\subsection{Creation of entanglement}
\begin{figure}[tbp]
\begin{center}
\includegraphics[width=6cm]{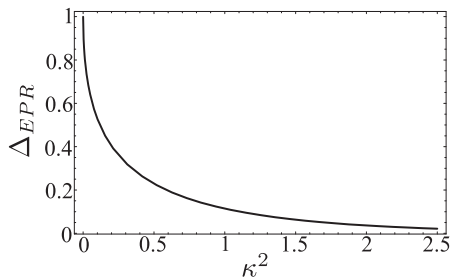}
\caption{Entanglement produced in the two mode squeezing scheme
versus coupling $\kappa^{2}$. The entanglement is hereby measured
by the EPR variance $\Delta_{EPR}$. \label{Entanglement}}
\end{center}
\end{figure}

As can be seen from the input-output relations for atoms and light
given in equations (\ref{atomic input-output relations squeezer})
and (\ref{light input-output relations squeezer}) respectively,
correlations between atoms and light are created which grow
exponentially in the coupling strength. We define new modes
appropriate to the type of correlations produced in the system by
setting
\begin{eqnarray*}
x_{1} &=& \frac{1}{\sqrt{2}}(x_{A}-\tilde{p}_{L+}),\ \ \ \ \ p_{1}
=
\frac{1}{\sqrt{2}}(p_{A}+\tilde{x}_{L+}),\\
x_{2}&=&\frac{1}{\sqrt{2}}(x_{A}+\tilde{p}_{L+}), \ \ \ \ \ p_{2}
= \frac{1}{\sqrt{2}}(p_{A}-\tilde{x}_{L+}).
\end{eqnarray*}
The corresponding variances can be calculated easily from
(\ref{atomic input-output relations squeezer}) and (\ref{light
input-output relations squeezer}). We get
\begin{eqnarray*}
(\Delta x_{1})^{2}&=&(\Delta
p_{2})^{2}=\left(\sqrt{e^{\kappa^{2}}-1}-e^{\frac{\kappa^{2}}{2}}\right)^{2}=e^{-2z},\\
(\Delta p_{1})^{2}&=&(\Delta
x_{2})^{2}=\left(\sqrt{e^{\kappa^{2}}-1}+e^{\frac{\kappa^{2}}{2}}\right)^{2}=e^{2z},\\
\end{eqnarray*}
with $z=\cosh^{-1}(e^{\frac{\kappa^{2}}{2}})$. We get a two mode
squeezed state where $x_{1}$ and $p_{2}$ are squeezed, while $p_{1}$
and $x_{2}$ are antisqueezed. In the limit of infinite coupling the
state becomes an EPR state in which $x_{A}$, $\tilde{p}_{L+}$ and
$p_{A}$, $\tilde{x}_{L+}$ are perfectly correlated. For the state
under consideration, the EPR-variance
$\Delta_{EPR}=\frac{1}{2}(\Delta x_{1}^{2}+\Delta
p_{2}^{2})=e^{-2z}$ is an entanglement measure \cite{Entanglement of
formation}. For separable states $\Delta_{EPR}=1$. For inseparable
states $\Delta_{EPR}$ decreases with increasing entanglement. The
amount of entanglement created in the scheme is shown in figure
\ref{Entanglement}.
\subsection{Spin squeezing}
\begin{figure}[tbp]
\begin{center}
\includegraphics[width=6cm]{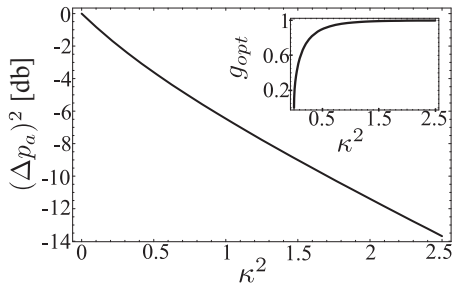}
\caption{Atomic squeezing $(\Delta p_{a})^{2}$ in db versus
coupling $\kappa^{2}$ for optimal gainfactor $g_{ opt}$. The inset
shows how the optimal gainfactor depends on the coupling.
\label{Squeezer}}
\end{center}
\end{figure}
The correlations created in the proposed scheme can be used to
produce atomic squeezing. This can be achieved by performing a
measurement on the plus light mode and subsequent feedback onto the
atomic spin based on the measurement outcome. The squeezing protocol
is symmetric with respect to the interchange of $ \{x_{A},\
\tilde{p}_{L}\}$ and $\{p_{A},\ \tilde{x}_{L}\}$. Here squeezing of
$(\Delta p_{A})^{2}$ is illustrated. In order to acquire information
about $p_{A}$, $\tilde{x}_{L+}$ has to be measured. The outcome of
this measurement is governed by the operator equation
\begin{eqnarray*}
\tilde{x}^{out}_{L+}=\sqrt{e^{\kappa^{2}}-1}\ p^{in}_{A}
+e^{\frac{\kappa^{2}}{2}}\tilde{x}^{in}_{L-}.
\end{eqnarray*}
If the measurement outcome $q_{L+}$ is obtained  $p^{out}_{A}$ is
displaced by an amount $g\ q_{L+}$, where $g \in \mathbb{R}$ is
some gain factor. For this feedback procedure the operator
identity
\begin{eqnarray*}
p^{fb}_{A}=p^{out}_{A}-g\  \tilde{x}^{out}_{L+}\
\end{eqnarray*}
holds in the ensemble average, as is shown in \cite{Teleportation,
JSCFP}. With help of the atomic input-output relations (\ref{atomic
input-output relations squeezer}) and the expression for the
measurement outcome above one finds
\begin{eqnarray*}
  p^{fb}_{A}\!= \!\!\bigg (\!e^{\frac{\kappa^{2}}{2}}-g
  \sqrt{e^{\kappa^{2}}-1}\bigg )p^{in}_{A}
  + \bigg (\!\! \sqrt{e^{\kappa^{2}}-1}-g e^{\frac{\kappa^{2}}{2}}
  \bigg )\tilde{x}^{in}_{L-}\ .
\end{eqnarray*}
Thus the variance of this quadrature is given by
\begin{eqnarray*}
(\Delta p^{fb}_{A})^{2}\!\!=\!\! \bigg
(\!e^{\frac{\kappa^{2}}{2}}-g
  \sqrt{e^{\kappa^{2}}-1}\!\bigg  )^{2}\!\frac{1}{2}
   +\! \bigg (\!\sqrt{e^{\kappa^{2}}-1}-ge^{\frac{\kappa^{2}}{2}}\!\bigg )^{2}
   \!\frac{1}{2}.
\end{eqnarray*}
$(\Delta p^{fb}_{A})^{2}$ is now optimized with respect to the
gainfactor $g$. We obtain
\begin{eqnarray*}
 g_{opt}&=& \frac{ e^{\frac{\kappa^{2}}{2}}\sqrt{e^{\kappa^{2}}-1}
 +e^{\kappa^{2}}\sqrt{1-e^{-\kappa^{2}}}}{2 e^{\kappa^{2}}-1}\ ,\\
( \Delta p^{fb}_{A \ opt})^{2}&=& \frac{1}{2} \ \frac{1}{2
e^{\kappa^{2}}-1}\ .
\end{eqnarray*}\\
Note that the atoms are left in a minimum uncertainty state, since
\begin{eqnarray*}
   (\Delta x_{A})^{2}=\frac{1}{2}(2
      e^{\kappa^{2}}-1)
   =\frac{1}{4}\ \frac{1}{(\Delta p^{fb}_{A})^{2}}\ .
\end{eqnarray*}
The amount of squeezing depending on the coupling $\kappa^{2}$ is
shown in figure \ref{Squeezer}.

\section{Consideration of noise}\label{Consideration of Noise}
We consider losses for both components of the protocol - atoms and
light - and treat them perturbativly within the Gaussian
formalism.
Concerning the atomic sample we take transverse decoherence of the
atomic spin state at a rate of $\frac{\eta}{T}$ into account.
As in experiments atomic vapor is usually contained within a glass
cell, the dominant source of noise concerning light are reflection
losses. These affect both, quantum variables and classical field and
will be characterized by the reflection coefficient $r$.
\subsection{Quantum memory with noise}\label{Quantum memory with
noise}
\begin{figure}[tbp]
\begin{center}
\includegraphics[width=8.9cm]{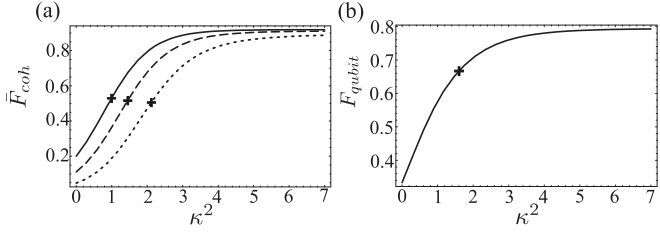}
\caption {Average fidelity with losses versus coupling. The atomic
decay rate $\eta$ and the reflection coefficient $r$ both have a
value of $7.5\%$. Crosses mark the corresponding classical limits.
(a) Fidelity for coherent input states according to distributions
with different mean photon numbers. (solid line: $n=4$, dashed
line: $n=8$, dotted line: $n=20$) (b) Fidelity for light qubits.
\label{FidelitiesWithNoise01}}
\end{center}
\end{figure}
\begin{figure}[tbp]
\begin{center}
\includegraphics[width=8.8cm]{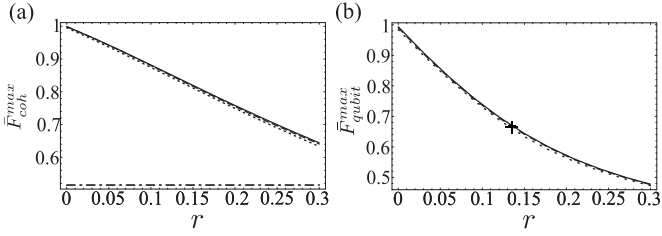}
\caption{Maximal attainable average fidelity for coherent input
states according to a distribution with mean photon number $n=8$
(a) and light qubits (b) versus reflection coefficient $r$ for
different atomic decay parameters $\eta$. (solid lines:
$\eta=5\%$, dashed lines: $\eta=10\%$, dotted lines: $\eta=25\%$)
The dash-dotted line and the cross indicate the classical limits.
\label{MaxFidelitiesWithNoise}}
\end{center}
\end{figure}
In this section we sum up results for write-in and read-out in the
presence of losses. A detailed description of the generalized
quantum memory scheme including noise is given in appendix
\ref{Consideration of noise in the memory protocol}.
Consideration of losses leads to a modification of the original
write-in mode. The generalized write-in quadratures preferred by
the system are given by
\begin{eqnarray*}
\left(%
\begin{array}{c}
 \!\! x^{in}\!\!\\
  \!\!p^{in} \!\!\\
\end{array}%
\right)\!\!\!\!&\propto&\!\! \!\int_{0}^{T}\!\!\!\!dt
e^{\frac{wt}{2}}
       \!R(t)\!\! \left[\! (1-r)\!\!
       \left(%
      \begin{array}{c}
         \!\!\bar{p}_{L}(ct,0) \!\! \\
       \!\! -\bar{x}_{L}(ct,0)\!\!  \\
      \end{array}%
      \right) \!\! +\!\! 2 r\!\! \left(%
 \begin{array}{c}
   \!\! \bar{p}_{L}(ct,0)\!\!\\
    \!\!\bar{x}_{L}(ct,0)\!\! \\
  \end{array}%
\right)\right]\!,
\end{eqnarray*}
where $w=\eta/T+\kappa^{2}(1-2r)/ T $. Both sources of noise -
reflection losses and spontaneous decay as well -  give rise to a
generalized exponent in the exponential modulation function. In
addition to the changed envelope, the light mode appearing in the
atomic input-output relation is further disturbed: it lies no longer
exactly at the upper sideband, but contains a small contribution
from the lower one, as can be seen by comparing the expression above
to (\ref{upper sideband}) and (\ref{lower sideband}). Since it is
experimentally advantageous to encode the input signal at sideband
modes, we define a generalized write-in mode (denoted by capital
letters)
\begin{eqnarray*}\nonumber
\left(%
\begin{array}{c}
  \!X_{us+}^{in}\! \\
  \!P_{us+}^{in}\!\\
\end{array}%
\right)\!\!&=&\!\!\sqrt{\frac{w}{e^{wT}-1}}\int_{0}^{T}dt\
e^{\frac{wt}{2}}
       R(t)\left(%
      \begin{array}{c}
         \!\bar{p}_{L}(ct,0) \! \\
         \!-\bar{x}_{L}(ct,0)\!  \\
      \end{array}%
      \right),
\end{eqnarray*}
which takes full account of noise concerning the exponential
modulation, but lies precisely at the upper sideband. (i.e. the
small orthogonal contribution from the lower one is treated as
noise.) To perform the read-out, the inverse accented counter part
of this mode is measured. The calculation of the fidelity is given
in appendix \ref{Fidelity for the complete state transfer
including noise}.
Figure \ref{FidelitiesWithNoise01} shows the average fidelity for
write-in and subsequent retrieval versus coupling for
$r=\eta=7.5\%$. Plots (a) and (b) refer to coherent input states
and light qubits respectively. The corresponding classical limits
are marked by crosses. As illustrated by these graphs losses
decrease not only the quality of the state transfer for a given
coupling strength, but limit also the attainable fidelity. The
crucial limiting factor in this scheme are reflection losses.
Figure \ref{MaxFidelitiesWithNoise} shows the maximum average
fidelity versus $r$ for different values of the atomic decay
parameter $\eta$. Plot (a) shows results for coherent inputs,
while plot (b) depicts the maximal attainable fidelity for qubits.
The dash-dotted line and the cross indicate the classical limits
in each case. Within moderate couplings fidelities well above the
classical limit can be achieved, showing that the protocol is
robust against the dominant sources of noise.
\subsection{Two mode squeezing with noise}
\begin{figure}[tbp]
\begin{center}
\includegraphics[width=6cm]{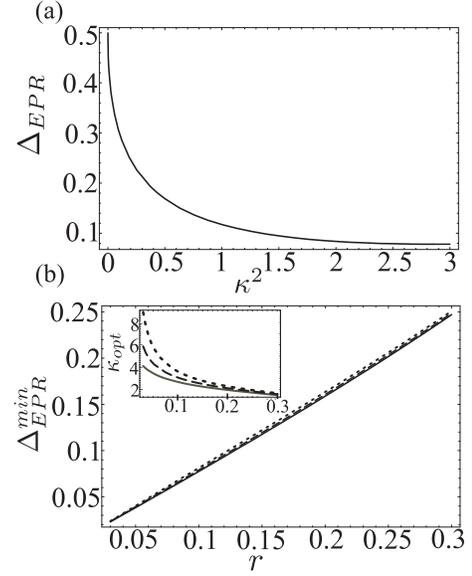} \caption{
\textbf{(a)} EPR variance $\Delta_{EPR}$ versus coupling
$\kappa^{2}$ in the presence of losses. The reflection coefficient
$r$ and the atomic decay rate are both chosen to have a value of
$10\%$.\\
\textbf{(b)} Optimized EPR variance versus coupling for different
atomic decay parameters. (solid line: $\eta= 5\%$, dashed line:
$\eta= 10\%$, dotted line: $\eta=25\%$) The inset shows how the
optimal coupling $k_{opt}$ varies with $r$. \label{EPRWithNoise}}
\end{center}
\end{figure}
\begin{figure}[tbp]
\begin{center}
\includegraphics[width=8.7cm]{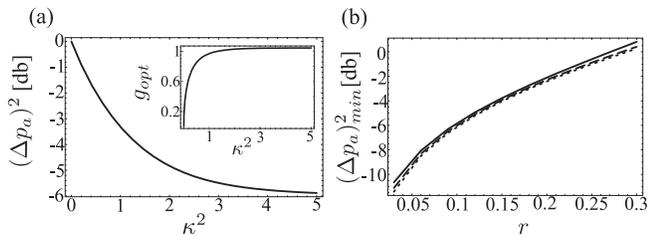} \caption{
\textbf{(a)} Spin squeezing in $db$ versus coupling $\kappa^{2}$
in the presence of losses. The reflection coefficient $r$ and the
atomic decay rate both have a value of $10\%$. The inset shows how
the optimal gainfactor $g_{opt}$ depends on the coupling.\\
\textbf{(b)} Maximal spin squeezing versus reflection coefficient
$r$ for different atomic decay parameters. (solid line: $\eta=
5\%$, dashed line: $\eta= 10\%$, dotted line: $\eta=25\%$).
\label{SqueezingWithNoise}}
\end{center}
\end{figure}

Consideration of noise within the two mode squeezing protocol is
done along the same lines outlined in the section above. The
entanglement created by the scheme in the presence of losses is
depicted in figure \ref{EPRWithNoise}(a) for $r=\eta=0.1$. The EPR
variance increases for higher values of $\kappa^{2}$. An optimal
value $\kappa_{opt}^{2}$ exists for which the proposed protocol
works best and a maximal amount of entanglement is generated.
Figure \ref{EPRWithNoise}(b) shows the $\kappa$-optimized EPR
variance versus $r$,while the dependence of $\kappa_{opt}$ on the
reflection coefficient is given within the inset. As can be seen
from these plots atomic decay plays a minor role.\\
Spin squeezing can be performed with a lower and limited quality in
the presence of losses. In contrast to the ideal case the optimal
gainfactor does not approach unity with increasing coupling but
converges towards a higher value which depends on the amount of
losses impairing the system. Figure \ref{SqueezingWithNoise}(a)
shows the squeezed atomic variance in $db$ and the dependence of
$g_{opt}$ on the coupling for $r=\eta=0.1$. The maximal attainable
squeezing versus $r$ is given in figure \ref{SqueezingWithNoise}(b)
for different atomic decay parameters.
\section{Conclusions}
In conclusion we propose two protocols based on a double-pass
scheme for a single atomic ensemble in a magnetic field. The first
protocol provides an exponential scaling interspecies
beam-splitter interaction. Therefore it is suitable for high
fidelity storage and retrieval of an unknown quantum state under
modest experimental conditions, as was shown for coherent input
states and light qubits as well. The second protocol generates
deterministically EPR entanglement between atoms and light. The
proposed protocols provide therefore the ingredients to realize a
variety of interesting quantum communication protocols. They are
also shown to remain experimentally feasible under realistic
conditions.
\subsection*{Acknowledgements}
We thank J. Sherson, A. Sorensen and K. Molmer for useful
discussions and acknowledge funding from the EU under project
FP6-511004-COVAQIAL, Integrated Project QAP and SCALA.

\appendix
\section{Consideration of noise in the memory protocol}\label{Consideration of noise in the memory protocol}
\subsubsection*{Write-in}
Atomic noise can be incorporated into the framework of section
\ref{Quantum memory} by including decay terms in Bloch equations
(\ref{bloch equation x memory}) and (\ref{bloch equation p
memory})
\begin{eqnarray}\label{generalized bloch equations memory}
  \partial_{t}x_{A}(t)\!\!\!&=&\!\!\!\Omega
    p_{A}(t)+\frac{\kappa}{\sqrt{T}}\overline{p}_{L}(ct,t)-\frac{\eta}{2 T} x_{A}(t) +\!\sqrt{\frac{\eta}{T}}f_{xA}(t),\nonumber\\
  \partial_{t}p_{A}(t)\!\!\!&=&\!\!\!-\!\Omega
    x_{A}(t)-\frac{\kappa}{\sqrt{T}}\overline{x}_{L}(ct-d,t) \nonumber\\
    &&-\frac{\eta}{2 T} p_{A}(t)
    +\!\sqrt{\frac{\eta}{T}}f_{pA}(t),
\end{eqnarray}
where $f_{xA}$ and $f_{pA}$ are Langevin noise operators with zero
mean and $\langle f(t)f(t')\rangle=\delta(t-t')\frac{1}{2}$.\\
\\Each time light crosses one of the cell walls, reflection losses
occur. This happens four times. In the following we will consider
coherent input states. In this case losses due to the very first
crossing can be neglected, since these can be compensated by using
a more intense input signal. In case of light qubit input states
losses due to the first reflection have to be considered, which
makes the resulting equations slightly more complicated. Since the
derivation is analogous except for this point it won't be sown
explicitly. Losses due to the second and third transit of a cell
wall affect only the second scattering interaction. We take this
into account by modifying the undisturbed equations for the light
field quadratures to be inserted into (\ref{generalized bloch
equations memory})
\begin{eqnarray*}
  \overline{x}_{L}(ct-d,t)&=& \overline{x}_{L}(ct-d,0)+\frac{\kappa}{\sqrt{T}}p_{A}(t-\frac{d}{c}),\\
  \overline{p}_{L}(ct,t)&=& \overline{p}_{L}(ct,0),
\end{eqnarray*}
by introducing light quadrature damping with a factor $2r$ (the
factor $2$ reflects the fact that crossing of a cell wall happens
twice) and corresponding light-Langevin operators $f_{xL}$ and
$f_{pL}$, and obtain
\begin{eqnarray*}
  \overline{x}_{L}(ct-d,t)&=&\sqrt{1-2r}\Big ( \overline{x}_{L}(ct-d,0)+\frac{\kappa}
    {\sqrt{T}}p_{A}(t-\frac{d}{c})\Big )\\
    &&+\sqrt{2r}f_{xL}(t),\\
  \overline{p}_{L}(ct,t)&=& \overline{p}_{L}(ct,0).
\end{eqnarray*}
$\overline{p}_{L}(ct,t)$ remains unchanged, since this quadrature
affects the atoms only in the first passage (during the
$p_{L}p_{A}\ $-interaction), which means that each part of the
pulse contributes before it is subjected to reflection losses.
Therefore $p_{L}$ is conserved as in the undisturbed case.
The classical light field is impaired by reflection losses as
well. Since the coupling strength of the scattering interaction is
proportional to the amplitude of the classical field we have a
reduced coupling $\tilde{\kappa}=\sqrt{1-2r}\ \kappa$ for the
second ($x_ {L}x_{A}$ -) interaction due to the light crossing two
cell walls before it's second passage. By considering this and
inserting the expressions above into equations (\ref{generalized
bloch equations memory}) we obtain
\begin{eqnarray*}
\partial_{t}x_{A}(t)\!\!&=&\!\!\Omega
p_{A}(t)\!+\!\frac{\kappa}{\sqrt{T}}\bar{p}_{L}(ct,0)\!-\!\frac{\eta}{2T}x_{A}(t)\!+\!\sqrt{\frac{\eta}{T}}f_{xA}(t),\\
\partial_{t}p_{A}(t)\!\!&=&\!\!-\Omega
x_{A}(t)\!-\!\frac{\tilde{\kappa}}{\sqrt{T}}\Big [\sqrt{1-2r}\Big
(\bar{x}_{L}(\xi,0)\!+\!\frac{\kappa}{\sqrt{T}}p_{A}(t)\Big)\\
&&\!+\!\sqrt{2r}f_{xL}(t)\Big
]\!-\!\frac{\eta}{2T}p_{A}(t)\!+\!\sqrt{\frac{\eta}{T}}f_{pA}(t).
\end{eqnarray*}
We can ignore reflection losses arising in the very last transit
through a cell wall, since the light field of the write-in beam is
of no relevance after the second scattering interaction. By
neglecting the time delay $d/c$ as in section \ref{Quantum
memory}, the atomic differential equations generalize to
\begin{eqnarray*}
   \partial_{t}\!\left(%
      \begin{array}{c}
        \!\!x_{A}(t)\!\! \\
       \! \!p_{A}(t)\!\! \\
      \end{array}%
      \right)\!\!\!&=&\!\! \Bigg \{\!\Omega\!\left(%
      \begin{array}{cc}
       \! \!\!0 & 1 \!\!\\
      \! \! \!-1 & 0\!\!\\
      \end{array}%
      \right)\!\!-\!\!\frac{\eta}{2T}\!\left(%
        \begin{array}{cc}
            \!\!1 & 0\!\! \\
            \!\!0 & 1 \!\!\\
          \end{array}%
       \right)\!-\frac{\kappa^{2}(1-2r)}{T}\!\left(%
       \begin{array}{cc}
       \! \!0 & 0 \!\!\\
        \!\!0 & 1\! \!\\
       \end{array}%
      \right)\!\!\!\Bigg \}\\
      &&\left(%
    \begin{array}{c}
     \!\!x_{A}(t)\!\! \\
     \!\!p_{A}(t)\! \!\\
   \end{array}%
   \right)\!+\frac{\kappa }{\sqrt{T}}\left(%
   \begin{array}{c}
         \bar{p}_{L}(ct,0)  \\
        \!-(1-2r)\bar{x}_{L}(ct,0)\!  \\
   \end{array}%
  \right)\\
  &&+\sqrt{\frac{\eta}{T}}\left(%
\begin{array}{c}
  \!\!f_{xA}(t)\! \!\\
  \!\!f_{pA}(t)\! \!\\
\end{array}%
\right)\!+\frac{\kappa \sqrt{2r}}{\sqrt{T}}\left(%
\begin{array}{c}
  0 \\
 \!\!\!-\sqrt{1-2r} f_{xL}(t)\!\!\!\\
\end{array}%
\right).
\end{eqnarray*}
We introduce the abbreviation $w=\eta/T+\kappa^{2}(1-2r)/T$, which
is the generalization of the exponent $\kappa^{2}/T$ of the
previous sections and change the previous assumption $2 \Omega T
\gg \kappa^{2}$ into $2 \Omega T \gg wT=\eta+\kappa^{2}(1-2r)$.
Therefore we get the homogeneous solution
$A(t)=e^{\frac{wt}{2}}R^{-1}(t)$, (where $R(t)$ is the rotation
matrix from section \ref{Quantum memory}) and thus
\begin{eqnarray}\label{generalized atomic time evolution}
     \left(%
        \begin{array}{c}
          \!x_{A}^{out}\! \\
          \!p_{A}^{out}\! \\
        \end{array}%
        \right)\!\!\!\!
 &=&\!\!e^{\frac{-wT}{2}}\!\left(%
       \begin{array}{c}
           \!x_{A}^{in}\!\\
           \!p_{A}^{in}\! \\
       \end{array}%
       \right)\\
       &&\!\!\!+e^{\frac{-wT}{2}}\
       \!\!\!\frac{\kappa}{\sqrt{T}}\!\int_{0}^{T}\!\!\!dte^{\frac{wt}{2}}
       R(t)\!\left(%
      \begin{array}{c}
         \bar{p}_{L}(ct,0)  \\
       \!\! -(1-2r)\bar{x}_{L}(ct,0)\!\!  \\
      \end{array}%
      \right) \nonumber \\
  \!\!\!\!&&\!\!\!+e^{\frac{-wT}{2}}\sqrt{\frac{\eta}{T}}
       \int_{0}^{T}\!\!\!dte^{\frac{wt}{2}}
       R(t)\!\left(%
      \begin{array}{c}
        f_{xA}(t) \\
        f_{pA}(t) \\
      \end{array}%
      \right) \nonumber \\
  \!\!\!&&\!\!\!+e^{\frac{-wT}{2}}\frac{\kappa \sqrt{2r}}{\sqrt{T}}
       \int_{0}^{T}\!\!\!dte^{\frac{wt}{2}}
       R(t)\!\left(%
       \begin{array}{c}
         \!\!0 \\
         \!\!-\sqrt{1-2r} f_{xL}(t)\!\!\\
      \end{array}%
     \right) \nonumber \!,
\end{eqnarray}
where $R(T)=\openone$ was used. The first two lines represent
atomic- and light contributions, while the third and fourth term
account for atomic noise and light noise respectively. The light
mode, which is naturally mapped onto the atomic sample, is no
longer a modulation of the upper sideband, as can be seen from the
factor $(1-2r)$ attached to $\bar{x}_{L}(ct,0)$ in the second
line. Since it is advantageous to encode the signal at sideband
modes, the term involving the new disturbed light mode is
decomposed into a generalization of the familiar plus mode
connected to the upper sideband
\begin{eqnarray}\nonumber
\left(%
\begin{array}{c}
  X_{us+}^{in} \\
  P_{us+}^{in} \\
\end{array}%
\right)&=&\sqrt{\frac{w}{e^{wT}-1}}\int_{0}^{T}dt\
e^{\frac{wt}{2}}
       R(t)\left(%
      \begin{array}{c}
         \bar{p}_{L}(ct,0)  \\
         -\bar{x}_{L}(ct,0)  \\
      \end{array}%
      \right)\\\label{generalized plus mode memory}
\end{eqnarray}
and a small contribution from an orthogonal plus mode lying at the
lower sideband
\begin{eqnarray*}
\left(%
\begin{array}{c}
  P_{ls+}^{in} \\
  X_{ls+}^{in} \\
\end{array}%
\right)&=&\sqrt{\frac{w}{e^{wT}-1}}\int_{0}^{T}dt\
e^{\frac{wt}{2}}
       R(t)\left(%
      \begin{array}{c}
         \bar{p}_{L}(ct,0)  \\
         \bar{x}_{L}(ct,0)  \\
      \end{array}%
      \right).
\end{eqnarray*}
Generalized light modes are denoted by capital letters. With this
decomposition the atomic input-output relations with noise read
\begin{eqnarray}\label{generalized atomic input-output relations memory}
     \left(%
        \begin{array}{c}
         \! \!x_{A}^{out}\!\! \\
         \! \!p_{A}^{out}\!\! \\
        \end{array}%
        \right)\!\!
 &=&\!\!e^{\frac{-wT}{2}}\!\left(%
       \begin{array}{c}
          \! \!x_{A}^{in}\!\!\\
           \!\!p_{A}^{in}\! \!\\
       \end{array}%
       \right)\\ \nonumber
       \!\!&&\!\!+\sqrt{1-e^{-wT}}\ \frac{\kappa(1-r)}{\sqrt{wT}}\left(%
\begin{array}{c}
 \!\! X^{in}_{us+}\!\! \\
 \!\! P^{in}_{us+}\!\! \\
\end{array}%
\right)\\ \nonumber
\!\!&&\!\!+\sqrt{1-e^{-wT}}\frac{\kappa r}{\sqrt{wT}} \left(%
\begin{array}{c}
  \!\!P^{in}_{ls+}\!\! \\
  \!\!X^{in}_{ls+}\!\! \\
\end{array}%
\right)\\ \nonumber
  \!\!&&\!\!+e^{\frac{-wT}{2}}\ \sqrt{\frac{\eta}{T}}
       \int_{0}^{T}\!\!dte^{\frac{wt}{2}}
       R(t)\!\!\left(%
      \begin{array}{c}
        \!\!f_{xA}(t)\!\! \\
        \!\!f_{pA}(t)\!\! \\
      \end{array}%
      \right)\\
  \!\!&&\!\!+e^{\frac{-wT}{2}}\frac{\kappa \sqrt{2r}}{\sqrt{T}}
       \int_{0}^{T}\!\!dte^{\frac{wt}{2}}
       R(t)\!\!\left(%
       \begin{array}{c}
         0 \\
         \!\!-\sqrt{1-2r} f_{xL}(t)\!\!\\
      \end{array}%
     \right)\!. \nonumber
\end{eqnarray}
\subsubsection*{Read-out}
In order to perform the read-out, a second pulse of light is sent
through the double pass scheme. Subsequently the light mode, which
is the inverse accented counter-part of the mode appearing in the
atomic time evolution (\ref{generalized atomic time evolution})
should be measured. Instead we choose the generalized minus mode
analogous to the write-in quadratures (\ref{generalized plus mode
memory}). The corresponding output quadratures are given by
\begin{eqnarray*}
\left(%
\begin{array}{c}
  \!\!\acute{X}_{us-}^{out}\!\! \\
  \!\!\acute{P}_{us-}^{out}\!\! \\
\end{array}%
\right)\!\!&=&\!\!\sqrt{\frac{w}{1-e^{-wT}}}\int_{0}^{T}dt\
e^{-\frac{wt}{2}}
       R(t)\left(%
      \begin{array}{c}
         \!\!\acute{\bar{p}}_{L}(ct,T)\!\!  \\
        \!\! -\acute{\bar{x}}_{L}(ct,T)\!\! \\
      \end{array}%
      \right).
\end{eqnarray*}
This can be evaluated by inserting the generalized expressions for
$\acute{\bar{p}}_{L}(ct,T)$ and $\acute{\bar{x}}_{L}(ct,T)$. For
$\acute{\bar{p}}_{L}(ct,T)$ we have
\begin{eqnarray*}
\acute{\bar{p}}_{L}(ct,T)&=&\sqrt{1-2r}\ \acute{\bar{p}}_{L}\
(ct,0)+\sqrt{2r}\acute{f}_{pL}(t)-\frac{\tilde{\kappa}}{\sqrt{T}}x_{A}(t).
\end{eqnarray*}
$\acute{\bar{p}}_{L}$ is damped after the first ($p$-conserving)
interaction and picks up some noise in return. Subsequently it
gets some $x_{A}$ -contribution during the second ($x_{L}x_{A}$-)
interaction. The reduced coupling strength $\tilde{\kappa}=
\sqrt{1-2r}\ \kappa$ accounts for the damped classical field in
the second passage. $\acute{\bar{x}}_{L}(ct,T)$ on the other hand
is given by
\begin{eqnarray*}
\acute{\bar{x}}_{L}(ct,T)\!\!&=&\!\! \sqrt{1-2r}\Big
(\acute{\bar{x}}_{L}(ct,0)+ \frac{\kappa}{\sqrt{T}}p_{A}(t)\Big
)+\sqrt{2r}\acute{f}_{xL}(t) .
\end{eqnarray*}
$\acute{\bar{x}}_{L}$ gets some $p_{A}$ contribution during the
first scattering interaction i.e. before the relevant transits
through cell walls occur. Subsequently this is damped and
appropriate noise is added.
All together both quadratures are damped, since both carry the
argument $(ct,T)$. This means each piece of the pulse contributes
after it ran trough the sample twice and has therefore already
experienced the two relevant transits trough cell walls.
The rest of the calculation is straight forward. In the end
reflection losses due to the fourth crossing of a cell wall have
to be considered by damping the calculated result by a factor
$\sqrt{1-r}$ and adding appropriate noise terms. The resulting
input-output relations for the read-out mode are
\begin{eqnarray}\nonumber\label{generalized light input-otput relations memory}
\left(%
\begin{array}{c}
  \acute{X}^{out}_{us-} \\
  \acute{P}^{out}_{us-} \\
\end{array}%
\right)\!\!&=&\!\! c_{1}
\left(%
\begin{array}{c}
  x^{in}_{A} \\
  p^{in}_{A}  \\
\end{array}%
\right) +c_{2}
\left(%
\begin{array}{c}
  \acute{X}_{us+}^{in} \\
  \acute{P}_{us+}^{in} \\
\end{array}%
\right)\nonumber \\
\!\!&&\!\!+c_{3}\left(%
\begin{array}{c}
   \acute{P}_{ls+}^{in} \\
   \acute{X}_{ls+}^{in} \\
\end{array}%
\right)
\!\!+c_{4}\left(%
\begin{array}{c}
  \acute{X}_{us-}^{in} \\
  \acute{P}_{us-}^{in} \\
\end{array}%
\right) \!\!+c_{5}
\left(%
\begin{array}{c}
   \acute{P}_{ls-}^{in} \\
   \acute{X}_{ls-}^{in} \\
\end{array}%
\right)
\nonumber  \\
&&\!\!+c_{6}\left(%
\begin{array}{c}
  F_{xA} \\
  F_{pA} \\
\end{array}%
\right)\!\!+ c_{7}\left(%
\begin{array}{c}
  \breve{F}_{xL} \\
  \breve{F}_{pL} \\
\end{array}%
\right)
\!\!+c_{8}\left(%
\begin{array}{c}
   \acute{F}_{xL} \\
   \acute{F}_{pL} \\
\end{array}%
\right)\nonumber \\
&&\!\!+c_{9}\!\!\int_{0}^{T} \!\!dt R(t)
[e^{-wT}e^{\frac{wt}{2}}\!\!-\!\!e^{-\frac{wt}{2}}]\left(%
\begin{array}{c}
  0 \\
 \!\! -\acute{f}_{xL}(t)\!\! \\
\end{array}%
\right).\nonumber \\
\end{eqnarray}
The coefficients $c_{1}$ to $c_{9}$ can  easily be calculated.
Since we want to focus on the structure of the equation, we don't
insert these prefactors in order to avoid complicated expressions.
The new read-out equations differ from (\ref{light input-output
relations memory}) by the appearance of noise terms (third and
fourth line) and extra light modes (second line). These
contributions are small and can be treated as perturbations.
$\left(%
\begin{array}{cc}
  F_{xA}, & F_{pA} \\
\end{array}%
\right)$
is an atomic noise mode, while
$\left(%
\begin{array}{cc}
  \breve{F}_{xL}, & \breve{F}_{pL} \\
\end{array}%
\right),$
refers to to the light mode which is due to the very last
reflection. It is independent from the light mode
$\left(%
\begin{array}{cc}
  \acute{F}_{pL}, & \acute{F}_{xL} \\
\end{array}%
\right)$
which accounts for the reflections happening between the
scattering interactions. These intermediate reflections give also
rise to terms in which only noise associated to $x_{L}$
contributes. They are summarized in the expression preceded by
$c_{8}$.
The appearance of light modes other than $\left(%
\begin{array}{cc}
  \acute{X}_{us+} & \acute{P}_{us+} \\
\end{array}%
\right)$ is due to a asymmetry between the $p_{L}p_{A}$
-interaction and the $x_{L}x_{A}$ present in a realistic setup in
contrast to the ideal case. The light field has to cross two glass
walls between the first and the second pass (thus affecting only
the $x_{L}x_{A}$ -interaction). Thus contributions from the lower
sideband appear  and contributions from the minus mode do not
cancel as in the ideal case.
\section{Fidelity for the complete state transfer including
noise}\label{Fidelity for the complete state transfer including
noise}
In the following subsections the fidelity for storage and
subsequent retrieval of an unknown state of light will be derived
for coherent input states and light qubits respectively. Hereby
reflection losses and transverse decoherence of the atomic spin
state are taken into account as explained in section V.A and
appendix A. The following calculations are based on the
input-output relations for the complete state transfer. For
coherent input states they are obtained by inserting equation
(\ref{generalized atomic input-output relations memory}), which
describes the atomic state after a noisy write-in procedure into
(\ref{generalized light input-otput relations memory}), which
gives us the final retrieved light state in the presence of
losses. For light qubit input states analogous relations hold.
Since we assume an atomic ensemble at room temperature, no
diffusion of the collective atomic mode during the storage time
has to be considered. Thermal motion of the atoms ensures, that
the collective mode which was addressed by the write-in beam is
identical with the collective atomic mode interacting with the
read-out pulse during the retrieval procedure. Decohering
mechanisms such as collisions impairing the atomic state during
the storage time occur on a slow scale. As was demonstrated in
\cite{JSCFP} storage times up to 10 ms can achieved.
\subsection*{Fidelity for coherent input states}
In order to compute the fidelity for coherent input states, means
and variances of the final quadratures have to be calculated.
$\langle\acute{X}_{L-}^{fin}\rangle,\langle\acute{P}_{L-}^{fin}\rangle$
and
$(\Delta\acute{P}_{L-}^{fin})^{2},(\Delta\acute{X}_{L-}^{fin})^{2}$
can be derived from the expression describing the complete state
transfer by using the assumption $2 \Omega T\gg wT=\eta
+\kappa^{2}(1-2r)$ (which is a direct generalization from the
approximation $2 \Omega T \gg \kappa ^{2}$ made in the ideal case)
and help of the noise operator properties $\langle
f_{x}\rangle=\langle f_{p}\rangle=\langle
f_{x}f_{p}+f_{p}f_{x}\rangle=0$ and $\langle
f(t)f(t')\rangle=\delta(t-t')\frac{1}{2}$. The obtained
expressions have to be inserted into equation (\ref{coherent
fidelity}), which gives the state overlap between the input-state
to be stored and the final state recieved. By considering a
gaussian distribution of width $n$ for coherent amplitudes the
average fidelity can be directly calculated as in section
\ref{Fidelity for coherent input states}.
\subsection*{Fidelity for light qubit input states including noise}
The initial qubit state $|\Psi_{in}\rangle=(\alpha+\beta a^{\dag
}_{in})|vac\rangle$ is subjected to the write-in and read-out
procedure which is represented by the unitary transformation
$U_{N}$. We obtain
\begin{eqnarray*}
|\Psi_{fin}\rangle&=&U_{N}|\Psi_{in}\rangle=(\alpha+\beta
U_{N}a^{\dag}_{in}U_{N}^{\dag})U_{N}|vac\rangle \\
&=&(\alpha+\beta a^{\dag}_{fin})U_{N}|vac\rangle.
\end{eqnarray*}
In contrast to the ideal case, where $U|vac\rangle=|vac\rangle$
could be used, $U_{N}$ is a general Bogoliubov transformation.
We remark that for $r=0$ the state transfer can still described by
a passive transformation. The active contribution is entirely due
to reflection losses. This can be understood, by noting that
reflection losses occurring between the first and the second
scattering interaction impair only the scattering in the second
pass. Therefore the active part of the second interaction cannot
compensate the active part in the first pass as in the ideal case.
This leads to a term in the generalized atomic input-output
relations, which contains only one light quadrature and can
therefore not be expressed as a mode-contribution. It plays an
isolated role in the commutation relations, but adds some extra
noise to the variances. Losses due to atomic decay on the other
hand are included into the
dynamics of the scheme in a symmetric way.\\
The fidelity for the complete state transfer is given by the
overlap between the target state
$|\Psi^{opt}_{fin}\rangle=(\alpha-\beta a^{\dag}_{in})|vac\rangle$
and the light state $|\Psi_{fin}\rangle$ which is effectively
retrieved
\begin{eqnarray}\label{generalized Qubit Fidelity 1}
F\!\!_{qubit}\!\!\!\!&=&\!\!\!\!
|\!\langle\Psi_{fin}^{opt}|\!\Psi_{fin}\rangle|^{2}\!\!=\!\!|\!\langle
vac|\!(\alpha^{*}\!\!\!-\!\!\beta^{*}\!a_{in})\!(\alpha\!+\!\beta
a^{\dag}_{fin}\!)U_{N}\!|vac\rangle|^{2}\!\!. \nonumber\\
\end{eqnarray}
$a^{\dag}_{fin}$ is known, since the input-output relations for
the complete state transfer are known. They can be written in
terms of creation and annihilation operators such that all
occurring modes are independent. The transformation is of the type
\begin{eqnarray} \label{generalized complete state transfer creation operators}
U_{N}a^{\dag}_{in}U_{N}^{\dag}=\sum_{i=1}^{n}k_{i}a_{i}^{\dag}+\sum_{j=1}^{m}\tilde{k}_{j}c_{j},
\end{eqnarray}
where the coefficients $k_ {i}$, $\tilde{k}_{j}$  are complex
numbers. $a^{\dag}_{1}=a^{\dag}_{in}$ refers to the state to be
stored, while $a_{2}^{\dag}$ to $a_{n}^{\dag}$ represent all
creation operators which appear in the equation, namely
contributions from the atomic input, atomic noise, light-input from
the read-out beam and light noise. Since we also have noise terms,
which cannot be expressed as a noise mode (compare equation
(\ref{generalized atomic input-output relations memory}) last term)
and contributions from the lower sideband (compare equation
(\ref{generalized light input-otput relations memory})) in which the
$x$- and $p$ quadratures are interchanged, we also have annihilation
operators in this equation which are represented by $c_{1}$ to
$c_{j}$. Since theses contributions are small, they are
treated as perturbations to the system.\\
The transformation given in (\ref{generalized complete state
transfer creation operators}) can be understood as an orthogonal
transformation $P=P_{a}\otimes P_{c}$, where $P_{a}$ acts on the
creation operators and $P_{c}$ acts on the annihilation operators,
followed by an active transformation $S$. With normalization
constants $N_{a}=\sqrt{\sum_{i=1}^{n}|k_{i}|^{2}}$ and
$N_{c}=\sqrt{\sum_{j=1}^{m}|\tilde{k}_{j}|^{2}}$, where
$N_{a}^{2}-N_{c}^{2}=1$ and $N_{c}\ll 1$, (\ref{generalized complete
state transfer creation operators}) can be written as
\begin{eqnarray}
a_{fin}^{\dag}&=&N_{a}\Big(\sum_{i=1}^{n}\frac{k_{i}}{N_{a}}a_{i}^{\dag}\Big)+N_{c}\Big(\sum_{j=1}^{m}\frac{\tilde{k}_{j}}{N_{c}}c_{j}\Big)\\
&=& N_{a}P_{a}a^{\dag}_{1}P_{a}^{\dag}+N_{c}P_{c}c_{1}P_{c}^{\dag}=N_{a}a^{\dag}_{P}+N_{c}c_{P}\nonumber\\
&=&\sqrt{1+N_{c}^{2}}a^{\dag}_{P}+N_{c}c_{P}=Sa^{\dag}_{P}S^{\dag}
\end{eqnarray}
and we have $U_{N}=S(P_{a}\otimes P_{c})$.
In order to compute $F_{qubit}$ from equation (\ref{generalized
Qubit Fidelity 1}) the expression $U_{N}|vac\rangle$ has to be
determined. $U_{N}|vac\rangle=S(P_{a}\otimes P_{c} )|vac\rangle=S
|vac\rangle$, since $P$ is a passive transformation. $S$ on the
other hand refers to a two mode squeezing operation. As mentioned
above active contributions are treated perturbatively. The
corresponding time evolution $S= e^ {N_{c}(
a_{P}c-a_{P}^{\dag}c^{\dag})}$ is expanded in a series to first
order and we obtain
\begin{eqnarray*}
U_{N}|vac\rangle&=&\frac{1}{\sqrt{1+|N_{c}|^{2}}}(1-N_{c}a_{P}^{\dag}c_{P}^{\dag})|vac\rangle
\end{eqnarray*}
By inserting this expression in equation (\ref{generalized Qubit
Fidelity 1}) and inserting the right hand side of
(\ref{generalized complete state transfer creation operators}) for
$a^{\dag}_{fin}$ the fidelity can be directly calculated. We find
\begin{eqnarray*}
F_{qubit}=\frac{1}{1+|N_{c}|^{2}}\Bigg(|\alpha|^{2}-|\beta|^{2}k_{1}\Bigg(1-\frac{|N_{c}|^{2}}{\sqrt{1+|N_{c}|^{2}}}\Bigg)\Bigg)^{2}.
\end{eqnarray*}
In order to obtain the average fidelity we set
$\alpha=\cos(\frac{\theta}{2})$ and
$\beta=\sin(\frac{\theta}{2})e^{i \phi}$ and integrate over the
whole Bloch sphere
$\bar{F}_{qubit}=\frac{1}{4 \pi}\int_{0}^{\pi}\int_{0}^{2 \pi}
F_{qubit}(\theta,\phi)\sin(\theta)d\theta d\phi$.
The results are shown in figures \ref{FidelitiesWithNoise01} and
\ref{MaxFidelitiesWithNoise}.
%
%
%

%
\end{document}